\documentclass[preprint]{aastex}

\begin{document}

\title{Large scale distribution of arrival directions of cosmic rays detected 
above $10^{18}$~eV at the Pierre Auger observatory}

\author{\textbf{The Pierre Auger Collaboration}$^{\dag}$ \\
P.~Abreu$^{63}$, M.~Aglietta$^{51}$, M.~Ahlers$^{94}$, E.J.~Ahn$^{81}$, I.F.M.~Albuquerque$^{15}$, D.~Allard$^{29}$, I.~Allekotte$^{1}$, J.~Allen$^{85}$, P.~Allison$^{87}$, A.~Almela$^{11,\: 7}$
% ...
\\ }
\affil{$^{\dag}$ Av. San Mart\'in Norte 306, 5613 Malarg\"ue, Mendoza, Argentina; www.auger.org \\
$^{1}$ Centro At\'{o}mico Bariloche and Instituto Balseiro (CNEA-UNCuyo-CONICET), San 
Carlos de Bariloche, 
Argentina \\
% ...
$^{7}$ Instituto de Tecnolog\'{\i}as en Detecci\'{o}n y Astropart\'{\i}culas (CNEA, CONICET, UNSAM), 
Buenos Aires, 
Argentina \\
% ...
$^{11}$ Universidad Tecnol\'{o}gica Nacional - Facultad Regional Buenos Aires, Buenos Aires,
Argentina \\
% ...
$^{15}$ Universidade de S\~{a}o Paulo, Instituto de F\'{\i}sica, S\~{a}o Paulo, SP, 
Brazil \\
% ...
$^{29}$ Laboratoire AstroParticule et Cosmologie (APC), Universit\'{e} Paris 7, CNRS-IN2P3, 
Paris, 
France \\
% ...
$^{51}$ Istituto di Fisica dello Spazio Interplanetario (INAF), Universit\`{a} di Torino and 
Sezione INFN, Torino, 
Italy \\
% ...
$^{63}$ LIP and Instituto Superior T\'{e}cnico, Technical University of Lisbon, 
Portugal \\
% ...
$^{81}$ Fermilab, Batavia, IL, 
USA \\
% ...
$^{85}$ New York University, New York, NY, 
USA \\
% ...
$^{87}$ Ohio State University, Columbus, OH, 
USA \\
% ...
$^{94}$ University of Wisconsin, Madison, WI, 
USA \\
% ...
(\ddag) Deceased \\
(a) at Konan University, Kobe, Japan \\
(b) now at the Universidad Autonoma de Chiapas on leave of absence from Cinvestav \\
(f) now at University of Maryland \\
(h) now at NYU Abu Dhabi \\
(i) now at Universit\'{e} de Lausanne \\
}

\begin{abstract}
A thorough search for large scale anisotropies in the distribution of arrival directions 
of cosmic rays detected above $10^{18}$~eV at the Pierre Auger Observatory is presented. 
This search is performed as a function of both declination and right ascension in several 
energy ranges above $10^{18}$~eV, and reported in terms of dipolar and quadrupolar 
coefficients. Within the systematic uncertainties, no significant deviation from isotropy 
is revealed. Assuming that any cosmic ray anisotropy is dominated by dipole and quadrupole 
moments in this energy range, upper limits on their amplitudes are derived. These upper 
limits allow us to challenge an origin of cosmic rays above $10^{18}$~eV from stationary 
galactic sources densely distributed in the galactic disk and emitting predominantly light 
particles in all directions.
\end{abstract}



\section{Introduction}

Establishing at which energy the intensity of extragalactic cosmic rays starts to dominate 
the intensity of galactic ones would constitute an important step forward to provide 
further understanding on the origin of Ultra-High Energy Cosmic Rays (UHECRs). 
A time honored picture is that the \textit{ankle}, a hardening of the energy spectrum 
located at $\simeq~$4~EeV~\citep{Linsley1963,Lawrence1991,Nagano1992,Bird1993,AugerPLB2010}
(where 1~EeV~$\equiv 10^{18}~$eV),
is the feature in the energy spectrum marking the transition between galactic and 
extragalactic UHECRs~\citep{Linsley1963}. As a natural signature of the escape of cosmic 
rays from the Galaxy, large scale anisotropies in the distribution of arrival directions could 
be detected at energies below this spectral feature. Both the amplitude and the shape of 
such patterns are uncertain, as they depend on the model adopted to describe the 
regular and turbulent components of the galactic magnetic field,  the charges of the cosmic 
rays, and the assumed distribution of sources in space and time. For cosmic rays mostly
heavy and originating from stationary sources located in the galactic disk, some estimates
based on diffusion and drift motions~\citep{Ptuskin1993,Candia2003} as well as direct 
integration of trajectories~\citep{Ptuskin1998,Giacinti2011} show that dipolar anisotropies 
at the level of a few percent could be imprinted in the energy range just below the ankle 
energy. Even larger amplitudes could result in the case of light primaries, unless sources are 
strongly intermittent and pure diffusion motions hold up to EeV energies~\citep{Calvez2010,Pohl2011}.

If UHECRs above 1~EeV have a predominant extragalactic 
origin~\citep{Hillas1967,Blumenthal1970,Berezinsky2006,Berezinsky2004}, their angular 
distribution is expected to be isotropic to a high level. But, even for isotropic extragalactic cosmic 
rays, the translational motion of the Galaxy relative to a possibly stationary extragalactic cosmic ray 
rest frame can produce a dipole in a similar way to the \textit{Compton-Getting effect}~\citep{Compton} 
which has been measured with cosmic rays of much lower energy at the solar time 
scale~\citep{Groom,Tibet,Milagro,EASTOP,IceCube} as a result of the Earth motion relative to the 
frame in which the cosmic rays have no bulk motion. Moreover, the rotation of the Galaxy can also 
produce anisotropy by virtue of moving magnetic fields, as cosmic rays travelling through far away 
regions of the Galaxy experience an electric force due to the relative motion of the system in which 
the field is purely magnetic~\citep{Harari2010}. The large scale structure of the galactic magnetic 
field is expected to transform even a simple Compton-Getting dipole into a more complex anisotropy 
at Earth, described by higher order multipoles~\citep{Harari2010}. A quantitative estimate of the 
imprinted pattern would require knowledge of the global structure of the galactic magnetic field and 
the charges of the particles, as well as the frame in which extragalactic cosmic rays have no bulk motion.
If, for instance, the frame in which the UHECR distribution is isotropic coincides with the cosmic 
microwave background rest frame, the amplitude of the simple Compton-Getting dipole would be about 
0.6\%~\citep{Kachelriess2006}. The same order of magnitude is expected if UHECRs have no bulk
motion with respect to the local group of galaxies. 

The large scale distribution of arrival directions of UHECRs as a function 
of the energy is thus one important observable to provide key elements 
for understanding their origin in the EeV energy range. 
Using the large amount of data collected by the Surface Detector (SD) array of the Pierre 
Auger Observatory, results of first harmonic analyses of the right ascension distribution 
performed in different energy ranges above $0.25~$EeV were recently 
reported~\citep{AugerAPP2011}. Upper limits on the dipole component in the equatorial 
plane were derived, being below 2\% at 99\% $C.L.$ for EeV energies and providing 
the most stringent bounds ever obtained. These analyses benefit from the almost uniform 
directional exposure in right ascension of the SD array of the Pierre Auger Observatory 
which is due to the Earth rotation, and they constitute a powerful tool for picking up any 
dipolar modulation in this coordinate. However, since this technique is not sensitive to a
dipolar component along the Earth rotation axis, we aim in the present 
report at estimating not only the dipole component in the right ascension distribution 
but also the component along the Earth rotation axis. More generally, we present 
a comprehensive search in all directions for any dipole or quadrupole patterns significantly 
standing out above the background noise. 

Searching for anisotropies with relative amplitudes down to the percent level requires
the control of the exposure of the experiment at even greater accuracy. Spurious 
modulations in the right ascension distribution are induced by the variations of the effective 
size of the SD array with time and by the variations of the counting rate of events due to 
the changes of atmospheric conditions. In Ref.~\citep{AugerAPP2011}, we showed in 
a quantitative way that such effects can be properly accounted for by making use of the 
instantaneous status of the SD array provided each second by the monitoring system, and 
by converting the observed signals in actual atmospheric conditions into the ones
that would have been measured at some given reference atmospheric conditions. 
Searching for anisotropies explicitly in declination requires the control of additional systematic 
errors affecting both the directional exposure of the Observatory and the counting rate of 
events in local angles. Each of these additional effects are carefully presented in 
sections~\ref{s1000} and~\ref{exposure}.

After correcting for the experimental effects, searches for large scale patterns 
above 1~EeV are presented in section~\ref{analysis}. Additional
cross-checks against eventual systematic errors affecting the results obtained in 
section~\ref{analysis} are presented in section~\ref{systematics}. 
Resulting upper limits on dipole and quadrupole amplitudes are presented and 
discussed in section~\ref{discussion}, while a final summary is given in 
section~\ref{summary}. Some further technical aspects are detailed in the
appendices.

\section{The Pierre Auger Observatory and the data set}
\label{pao}

The Pierre Auger Observatory~\citep{AugerNIM2004}, located in Malarg\"{u}e, Argentina, 
at mean latitude 35.2$^\circ\,$S, mean longitude 69.5$^\circ\,$W and mean altitude 1400 
meters above sea level, has been designed to collect UHECRs with unprecedented
statistics. It exploits two available techniques to detect extensive air showers initiated by
cosmic ray interactions in the atmosphere~: a \textit{surface detector array} and 
a \textit{fluorescence detector}. The SD array consists of 1660 water-Cherenkov detectors 
laid out over about 3000~km$^2$ on a triangular grid with 1.5~km spacing. These 
water-Cherenkov detectors are sensitive to the light emitted in their volume by the
secondary particles of the showers, and provide a lateral sampling of the showers reaching
the ground level. At the perimeter of this array, the atmosphere is overlooked on dark 
nights by 27 optical telescopes grouped in 5 buildings.
These telescopes record the number of secondary charged particles in the air shower as a 
function of depth in the atmosphere by measuring the amount of nitrogen fluorescence caused 
by those particles along the track of the shower.

The analyses presented in this report make use of events recorded by the SD array from 1 January 
2004 to 31 December 2011, with zenith angles less than 55$^\circ$. To ensure good angle and
energy reconstructions, each event must satisfy a fiducial cut requiring that the \emph{elemental cell}
of the event (that is, the all six neighbours of the water-Cherenkov detector with the highest signal)
was active when the event was recorded~\citep{AugerNIM2010}. Based on this fiducial cut, and 
accounting for unavoidable periods of array instability reducing slightly the duty cycle, 
the total geometric exposure corresponding to the data set considered in this report is 
23,520~km$^2$~yr~sr. This geometric exposure applies to energies at which the SD array
operates with full detection efficiency, that is, to energies above 3~EeV~\citep{AugerNIM2010}.

The event direction is determined following the procedure described in Ref.~\citep{ang-res}. At the lowest
energies observed, the angular resolution of the SD is about $2.2^\circ$, and reaches $\sim 1^\circ$ at the
highest energies~\citep{ang-res2}. This is sufficient to perform searches for large-scale anisotropies.

The energy estimation of each event is primarily based on the measurement of the 
signal at a reference distance of $1000\,$m, $S(1000)$, referred to as the \textit{shower size}.
For a given energy, the shower size is a function of the zenith angle due to the rapid increase 
of the slant depth which induces an attenuation of the electromagnetic component of the
showers. To account for this attenuation, the relationship between the observed $S(1000)$
and the one that would have been measured had the shower arrived at a zenith angle 38$^\circ$
is derived in an empirical way, using the constant intensity cut method~\citep{Hersil1961}.
To convert $S_{38^\circ}$ into energy, a calibration curve is used, based on events measured 
simultaneously by the SD array and the fluorescence telescopes~\citep{AugerPRL2008},
since these telescopes indeed provide a calorimetric measurement of the energy.
The statistical uncertainty of this energy estimation amounts to about 15\%, while the absolute 
energy scale has a systematic uncertainty of 22\%~\citep{AugerPRL2008}.

\section{Control of the event counting rate}
\label{s1000}

The control of the event counting rate is critical in searches for large scale \linebreak  anisotropies.
Due to the steepness of the energy spectrum, any mild bias in the estimate of the shower
energy with time or incident angles can lead to significant distortions of the event counting rate.
The procedure followed to obtain an unbiased estimate of the shower energy is described
in this section. This procedure consists in correcting measurements of shower sizes, $S(1000)$, 
for the influences of weather effects and the geomagnetic field \emph{before} 
the conversion to $S_{38^\circ}$ using the constant intensity method. Then, the conversion
to energy is applied. 

\subsection{Influence of atmospheric conditions on shower size}
\label{s1000-w}

\begin{table*}[h]
\begin{center}
\begin{tabular}{c|c|c|c}
$\sec{\theta}$ & $\alpha_\rho$[kg$^{-1}$m$^3$] & $\beta_\rho$[kg$^{-1}$m$^3$] & $\alpha_P$[hPa$^{-1}$] \\
\hline
\hline
$[1.0 - 1.2]$ & $-9.7~10^{-1}$ & $-2.6~10^{-1}$ & $-4.4~10^{-4}$  \\
$[1.2 - 1.4]$ & $-7.2~10^{-1}$ & $-2.2~10^{-1}$ & $-1.6~10^{-3}$  \\
$[1.4 - 1.6]$ & $-5.4~10^{-1}$ & $-2.0~10^{-1}$ & $-2.3~10^{-3}$  \\
$[1.6 - 1.8]$ & $-4.0~10^{-1}$ & $-4.3~10^{-2}$ & $-1.9~10^{-3}$  \\
$[1.8 - 2.0]$ & $-1.5~10^{-1}$ & $-2.3~10^{-2}$ & $-2.8~10^{-3}$ 
\end{tabular}
\caption{\small{Coefficients $\alpha_\rho$, $\beta_\rho$ and $\alpha_P$ used to correct 
shower sizes for atmospheric effects on shower development, in bins of $\sec{\theta}$. 
From Ref.~\citep{AugerAPP2009}.}}
\label{tab:weather}
\end{center}
\end{table*}%
The energy estimator of the showers recorded by the SD array is provided by the signal 
at 1000~m from the shower core, $S(1000)$. For any fixed energy, since the development
of extensive air showers depends on the atmospheric pressure $P$ and air density $\rho$,
the corresponding $S(1000)$ is sensitive to variations in pressure and air density. Systematic
variations with time of $S(1000)$ induce variations of the event rate that may distort the
real dependence of the cosmic ray intensity with right ascension. To cope with this experimental
effect, the observed shower size $S(1000)$, measured at the actual density $\rho$ and pressure $P$, 
is related to the one $S_{atm}(1000)$ that would have been measured at reference values 
$\rho_0$ and $P_0$~\citep{AugerAPP2009}~:
\begin{equation}
S_{atm}(1000)=\left[1-\alpha_P(\theta)(P-P_0)-\alpha_\rho(\theta)(\rho_d-\rho_0)-\beta_\rho(\theta)(\rho-\rho_d)\right]S(1000).
\label{sweather}
\end{equation}
The reference values are chosen as the average values at Malarg\"ue (\textit{i.e.} 
$\rho_0=1.06$ kg~m$^{-3}$ and $P_0=862$~hPa). $\rho_d$ denotes here the average daily density 
at the time the event was recorded. The measured coefficients $\alpha_\rho$, $\beta_\rho$ and 
$\alpha_P$ - given in Table~\ref{tab:weather} - give the influence on the shower sizes of the
air density (and thus temperature) at long and short time scales on the Moli\`ere radius (and hence
the lateral profiles of the showers) and of the pressure on the longitudinal development of air showers, 
respectively. 

Applying these corrections to the energy assignments of showers allows us to cancel spurious
variations of the event rate in right ascension, whose typical amplitudes amount to a few per thousand
when considering data sets collected over full years.

\subsection{Influence of the geomagnetic field on shower size}
\label{s1000-g}

The trajectories of charged particles in extensive air showers are curved in the Earth's magnetic
field, resulting in a broadening of the spatial distribution of particles in the direction of the 
Lorentz force. As the strength of the geomagnetic field component perpendicular to any
arrival direction depends on both the zenith and azimuthal angles, the small changes of
the density of particles at ground induced by the field break the circular symmetry of the
lateral spread of the particles and thus induce a dependence of the shower size
$S(1000)$ at a fixed energy in terms of the azimuthal angle. Due to the steepness of the
energy spectrum, such an azimuthal dependence translates into azimuthal modulations of 
the estimated cosmic ray event rate at a given $S(1000)$. To eliminate these effects, the observed
shower size $S(1000)$ is related to the one that would have been observed in the absence
of geomagnetic field $S_{geom}(1000)$~\citep{AugerJCAP2011}~:
\begin{eqnarray}
S_{geom}(1000)=\left[1-g_1\cos^{-g_2}{(\theta)}\sin^2{(\widehat{\textbf{u},\textbf{b}})}\right]S(1000),
\label{sgeom}
\end{eqnarray}
where $g_1=(4.2\pm1)~10^{-3}$, $g_2=2.8\pm0.3$, and 
$\textbf{u}$ and $\textbf{b}=\textbf{B}/\Vert \textbf{B} \Vert$ denote the unit vectors in the 
shower direction and the geomagnetic field direction, respectively. At a zenith angle $\theta=55^\circ$,
the amplitude of the asymmetry in azimuth already amounts to $\simeq2\%$, which is why we restrict 
the present analysis to zenith angles smaller than this value. Carrying out these corrections
is thus critical for performing large scale anisotropy measurements in declination.

\subsection{From shower size to energy}
\label{s1000-cic}

Once the influence on $S(1000)$ of weather and geomagnetic effects are accounted for, 
the dependence of $S(1000)$ on zenith angle due to the attenuation of the shower and 
geometrical effects is extracted from the data using the constant intensity cut 
method~\citep{AugerPRL2008}. The attenuation curve $CIC(\theta)$ is fitted with a second 
order polynomial in $x=\cos^2{(\theta)}-\cos^2{(38^\circ)}$~: $CIC(\theta)=1+ax+bx^2$. 
The angle $38^\circ$ is chosen as a reference to convert $S(1000)$ to 
$S_{38^\circ}=S(1000)/CIC(\theta)$. $S_{38^\circ}$ may be regarded as the signal that 
would have been expected had the shower arrived at $38^\circ$. The values of the parameters 
$a=0.94\pm 0.03$ and $b=-0.95\pm 0.05$ are deduced for 
$S_{38^\circ}=22~$VEM\footnote{A vertical equivalent muon, or VEM, is the expected signal 
in a surface detector crossed by a muon traveling vertically and centrally to it.}, that corresponds to an 
energy of about 4~EeV - just above the threshold energy for full efficiency. The differences of 
these parameters with respect to previous reports will be discussed in section~\ref{systematics}.

Finally, the sub-sample of events recorded by both the fluorescence telescopes and the
SD array is used to establish the relationship between the energy reconstructed with
the fluorescence telescopes $E_{FD}$ and $S_{38^\circ}$~: $E_{FD}=AS_{38^\circ}^B$. The 
resulting parameters from the data fit are 
$A=(1.68\pm0.05)\times10^{-1}~$EeV and $B=1.030\pm0.009$, 
in good agreement with the recent report given in Ref.~\citep{AugerPesce}. The energy scale 
inferred from this data sample is applied to all showers detected by the SD array.

\section{Directional exposure of the Surface Detector array above 1~EeV}
\label{exposure}

The \textit{directional exposure} $\omega$ of the Observatory provides the effective 
time-integrated collecting area for a flux from each direction of the sky~\footnote{In 
other contexts such as the determination of the energy spectrum 
for instance, the term "exposure" refers to the \emph{total} exposure integrated over the 
celestial sphere, in units km$^2$~yr~sr.}, in units km$^2$~yr. For energies below 3~EeV, it is controlled 
by the detection efficiency $\epsilon$ for triggering. This efficiency depends on the energy $E$, 
the zenith angle $\theta$, and the azimuth angle $\varphi$. Consequently, the directional exposure 
of the Observatory is maximal above 3~EeV, and it is smaller at lower energies where the detection 
efficiency is less than unity. 

In this section we show in a comprehensive way how the directional exposure of the SD array is
obtained as a function of the energy. We first explain how the slightly non-uniform exposure of the 
sky in sidereal time can be accounted for in the search for anisotropies (section~\ref{sub:exposure}). 
In section~\ref{sub:efficiency} we empirically calculate the detection efficiency as a function of 
the zenith angle and deduce the exposure below the full efficiency energy (3~EeV). In 
section~\ref{sub:geomagnetic} we discuss the azimuthal dependence of the efficiency due to the 
geomagnetic effects, introduce the corrections due to the tilt of the array in section~\ref{sub:tilt} 
and the corrections due to the spatial extension of the array in section~\ref{sub:extension} and
show that the influence of weather effects is negligible on the detection efficiency between 1 and 3~EeV
in section~\ref{sub:other}. Finally we give in section~\ref{sub:examples} some examples of our fully 
corrected exposure at several energies.

\subsection{From local to celestial directional exposure.}
\label{sub:exposure}

The choice of the fiducial cut to select high quality events 
allows the precise determination of the geometric directional aperture per cell as 
$a_{\mathrm{cell}}(\theta)=1.95~\cos{\theta}~$km$^2$~\citep{AugerNIM2010}. It also allows 
us to exploit the regularity of the array for obtaining its geometric directional aperture as a 
simple multiple of $a_{\mathrm{cell}}(\theta)$~\citep{AugerNIM2010}. The number of elemental 
cells $n_{\mathrm{cell}}(t)$ is accurately monitored every second at the Observatory. To 
search for celestial large scale anisotropies, it is mandatory to account for the modulation 
imprinted by the variations of $n_{\mathrm{cell}}(t)$ in the expected number of events at the 
\emph{sidereal periodicity} $T_{sid}$. Within each sidereal day, and in the same way as
in Ref.~\citep{AugerAPP2011}, we denote by $\alpha^0$ the local sidereal time and express 
it in hours or in radians, as appropriate. For practical reasons, $\alpha^0$ is chosen so that it is 
always equal to the right ascension of the zenith at the centre of the array. As a function of $\alpha^0$, 
the total number of elemental cells $N_{\mathrm{cell}}(\alpha^0)$ and its associated relative variations 
$\Delta N_{\mathrm{cell}}(\alpha^0)$ are then obtained from~:
\begin{equation}
\label{Ncell}
N_{\mathrm{cell}}(\alpha^0)=\sum_{j}n_{\mathrm{cell}}(\alpha^0+jT_{sid}), \hspace{1cm} \Delta N_{\mathrm{cell}}(\alpha^0)=\frac{N_{\mathrm{cell}}(\alpha^0)}{\left<N_{\mathrm{cell}}\right>_{\alpha^0}},
\end{equation}
with $\left<N_{\mathrm{cell}}\right>_{\alpha^0}=1/T_{sid}\int_0^{T_{sid}}\mathrm{d}\alpha^0N_{\mathrm{cell}}(\alpha^0)$.
In the same way as in Ref.~\citep{AugerAPP2011}, the small modulation of the expected number
of events in right ascension induced by those variations will be accounted for by  weighting each 
event $k$ with a factor inversely proportional to $\Delta N_{\mathrm{cell}}(\alpha^0_k)$ when 
estimating the anisotropy parameters in section~\ref{analysis}. Placing such time dependences
in the event weights allows us to remove the modulations in time imprinted by the growth of the 
array and the dead times for each detector.

At any time, the \textit{effective} directional aperture of the SD array is controlled by the 
geometric one \textit{and} by the detection efficiency function $\epsilon(\theta,\varphi,E)$. 
For each elemental cell, the directional exposure in celestial coordinates is then simply obtained 
through the integration over local sidereal time of 
$x^{(i)}(\alpha^0)\times a_{\mathrm{cell}}{(\theta)}\times\epsilon(\theta,\varphi,E)$, where 
$x^{(i)}(\alpha^0)$ is the operational time of the cell $(i)$. Actually, since the small modulations
in time imprinted in the event counting rate by experimental effects will be accounted for
by means of the weighting procedure just described when searching for anisotropies, 
the small variations in local sidereal time for each $x^{(i)}(\alpha^0)$ can be neglected in
calculating $\omega$. The zenith and azimuth angles are related to the declination and 
the right ascension through~:
\begin{eqnarray}
\label{eqn:theta-phi}
\cos{\theta}&=&\sin{\delta}\sin{\ell_\mathrm{site}}+\cos{\delta}\cos{\ell_\mathrm{site}}\cos{(\alpha-\alpha^0)},\nonumber\\
\tan{\varphi}&=&\frac{\cos{\delta}\sin{\ell_\mathrm{site}}\cos{(\alpha-\alpha^0)}-\sin{\delta}\cos{\ell_\mathrm{site}}}{\cos{\delta}\sin{(\alpha-\alpha^0)}},
\end{eqnarray}
with $\ell_{\mathrm{site}}$ the mean latitude of the Observatory.
Since both $\theta$ and $\varphi$ depend only on the difference $\alpha-\alpha^0$, the integration 
over $\alpha^0$ can then be substituted for an integration over the hour angle $\alpha^\prime=\alpha-\alpha^0$ 
so that the directional exposure actually does not depend on right ascension when the $x^{(i)}$
are assumed local sidereal time independent~:
\begin{equation}
\label{eqn:omega}
\omega(\delta,E) = \sum_{i=1}^{n_{\mathrm{cell}}}x^{(i)}\int_0^{24h}~\mathrm{d}\alpha^\prime\,a_{\mathrm{cell}}{(\theta(\alpha^\prime,\delta))}~\epsilon(\theta(\alpha^\prime,\delta),\varphi(\alpha^\prime,\delta),E).
\end{equation}
Above 3~EeV, this integration can be performed analytically~\citep{Sommers2001}. 
Below 3~EeV, the non-saturation of the detection efficiency makes the directional exposure lower.
The next sections are dedicated to the determination of $\epsilon(\theta,\varphi,E)$.

\subsection{Detection efficiency}
\label{sub:efficiency}

To determine the detection efficiency function, a natural method would be to generate showers by 
means of Monte-Carlo simulations and to calculate the ratio of the number of triggered events to 
the total simulated. However, there are discrepancies in the predictions of the hadronic 
interaction model regarding the number of muons in shower simulations and what is found in our
data~\citep{Engel2007}. This prevents us from relying on this method for obtaining the detection efficiency 
to the required accuracy. 

We adopt here instead an empirical approach, based on the quasi-invariance of the zenithal 
distribution to large scale anisotropies for zenith angles less than $\simeq 60^\circ$ and for 
any Observatory whose latitude is far from the poles of the Earth. For full efficiency, the distribution in 
zenith angles $\mathrm{d}N/\mathrm{d}\theta$ is proportional to $\sin{\theta}\cos{\theta}$ for 
solid angle and geometry reasons, so that the distribution in $\mathrm{d}N/\mathrm{d}\sin^2{\theta}$ 
is uniform. Consequently, below full efficiency, \emph{any significant deviation from a uniform 
behaviour in the $\mathrm{d}N/\mathrm{d}\sin^2{\theta}$ distribution provides an empirical 
measurement of the zenithal dependence of the detection efficiency}. The quasi-invariance of 
$\mathrm{d}N/\mathrm{d}\sin^2{\theta}$ to large scale anisotropies is demonstrated in Appendix A. 

\begin{figure}[!t]
  \centering					 
  \includegraphics[width=10cm]{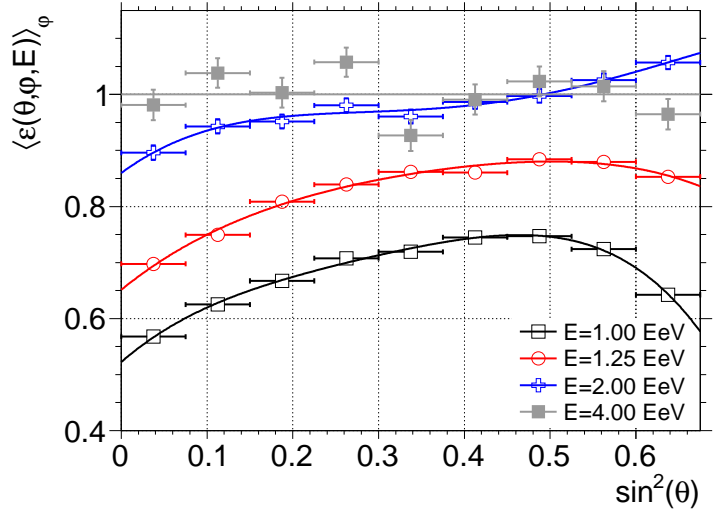}
  \caption{\small{Detection efficiency averaged over the azimuth as a function of $\sin^2{\theta}$
  at different energies, empirically measured from the data.}}
\label{fig:deteff-th}
\end{figure}

Based on this quasi-invariance, the detection efficiency averaged over the azimuth can be estimated from~:
\begin{equation}
\label{eqn:eps1}
\left<\epsilon(\theta,\varphi,E)\right>_{\varphi} = \frac{1}{\mathcal{N}}\frac{\mathrm{d}N(\sin^2{\theta},E)}{\mathrm{d}\sin^2{\theta}},
\end{equation}
where the notation $\left<\cdot\right>_{\varphi}$ stands for the average over $\varphi$ and
the constant $\mathcal{N}$ is the number of events that would have been observed at
energy $E$ and for any $\sin^2{\theta}$ value in case of full efficiency for an energy spectrum 
$\mathrm{d}N/\mathrm{d}E=40~(E/\mathrm{EeV})^{-3.27}~$km$^{-2}$yr$^{-1}$sr$^{-1}$EeV$^{-1}$ - 
as measured between 1 and 4~EeV~\citep{AugerPLB2010}. Consequently, for each zenith angle,
this empirical measurement of the efficiency provides an estimate \textit{relative} to the
overall spectrum of cosmic rays. In particular, since it is applied to \textit{all} events detected
at energy $E$ without distinction based on the primary mass of cosmic rays, this technique does 
not provide the mass dependence of the detection efficiency. For that reason, the
anisotropy searches reported in section~\ref{analysis} pertain to the whole population of 
cosmic rays, whether this population consists of a single primary mass or a mixture of several
elements.

Results are shown in Fig.~\ref{fig:deteff-th} for four different energies\footnote{To get the detection 
efficiency at a single energy $E$, events are actually selected in narrow energy bins around $E$. 
In addition, to account for the energy spectrum in $E^{-3.27}$ in this energy range, each event is 
weighted by a factor $E^{3.27}$.}. At 4~EeV, a uniform behaviour around 1 is observed, though 
quite noisy due to the reduced statistics. This uniform behaviour is consistent with full efficiency 
at this energy, as expected. Note that some values are greater than 1 for energies close or higher 
than 3~EeV, because of the empirical way of measuring the efficiency relative to the overall 
spectrum of cosmic rays. At 2~EeV, a loss of efficiency is 
observed for vertical showers due to the attenuation of the electromagnetic component of the 
showers. Up to $\simeq 40^\circ$, the detection efficiency steadily increases because the projected 
area of showers at ground gets larger with zenith angle. Above $\simeq 40^\circ$, the rapid increase 
of the slant depth makes then the attenuation of the electromagnetic component stronger, but the 
muonic component of showers becomes dominant and ensures a high detection efficiency. At lower 
energies, the number of muons is, in contrast, too low to impact significantly on the detection efficiency 
above $\simeq 40^\circ-45^\circ$, so that a clear decrease is observed at high zenith angles. In the following, 
we use parameterisations obtained by fitting each distribution with a fourth-order polynomial function 
in $\sin^2{\theta}$, which is sufficient to reproduce the main details as illustrated in Fig.~\ref{fig:deteff-th}.

\subsection{Geomagnetic effects below full efficiency}
\label{sub:geomagnetic}

\begin{figure}[!t]
  \centering					 
  \includegraphics[width=7.5cm]{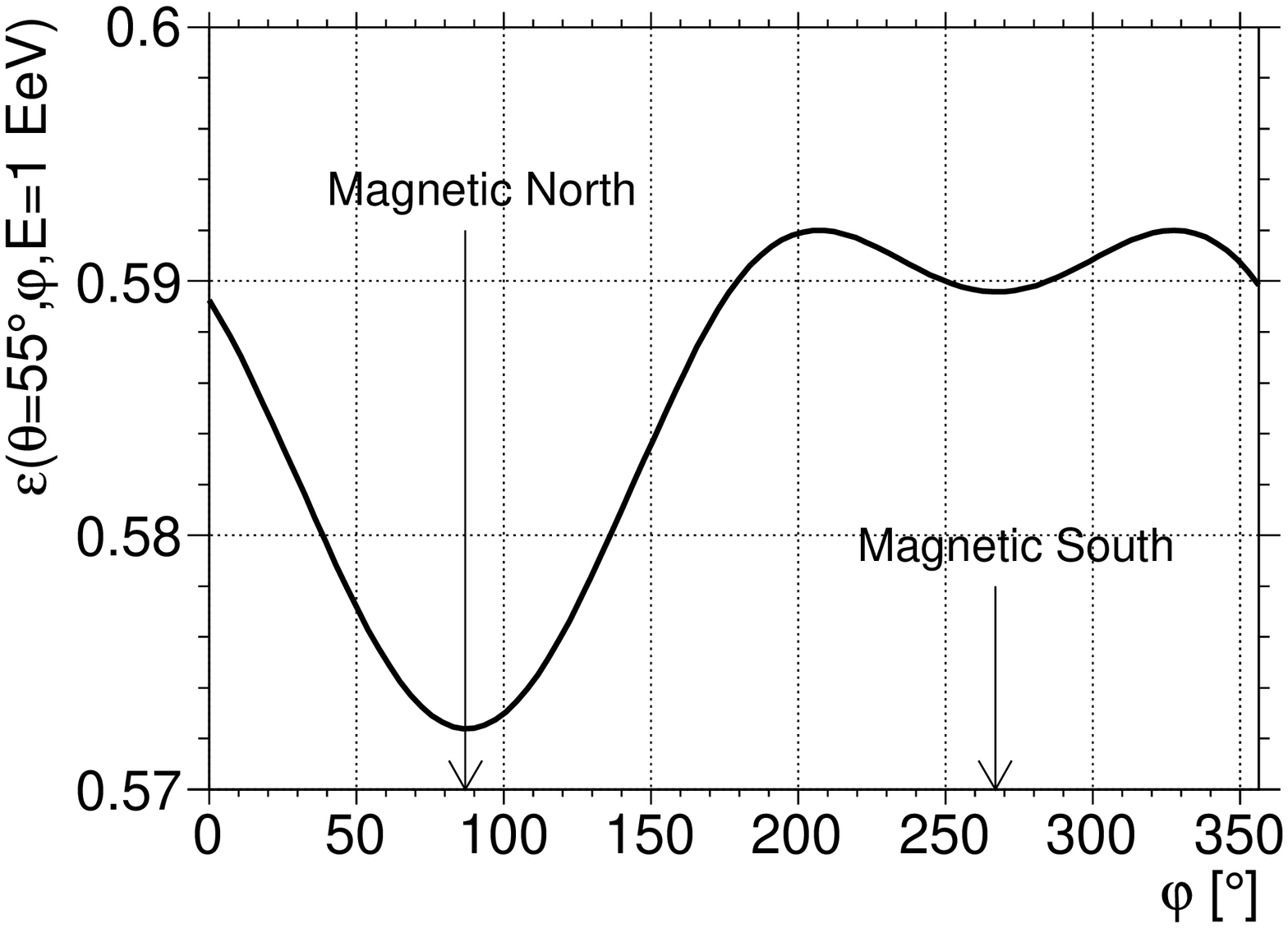}
  \includegraphics[width=7.5cm]{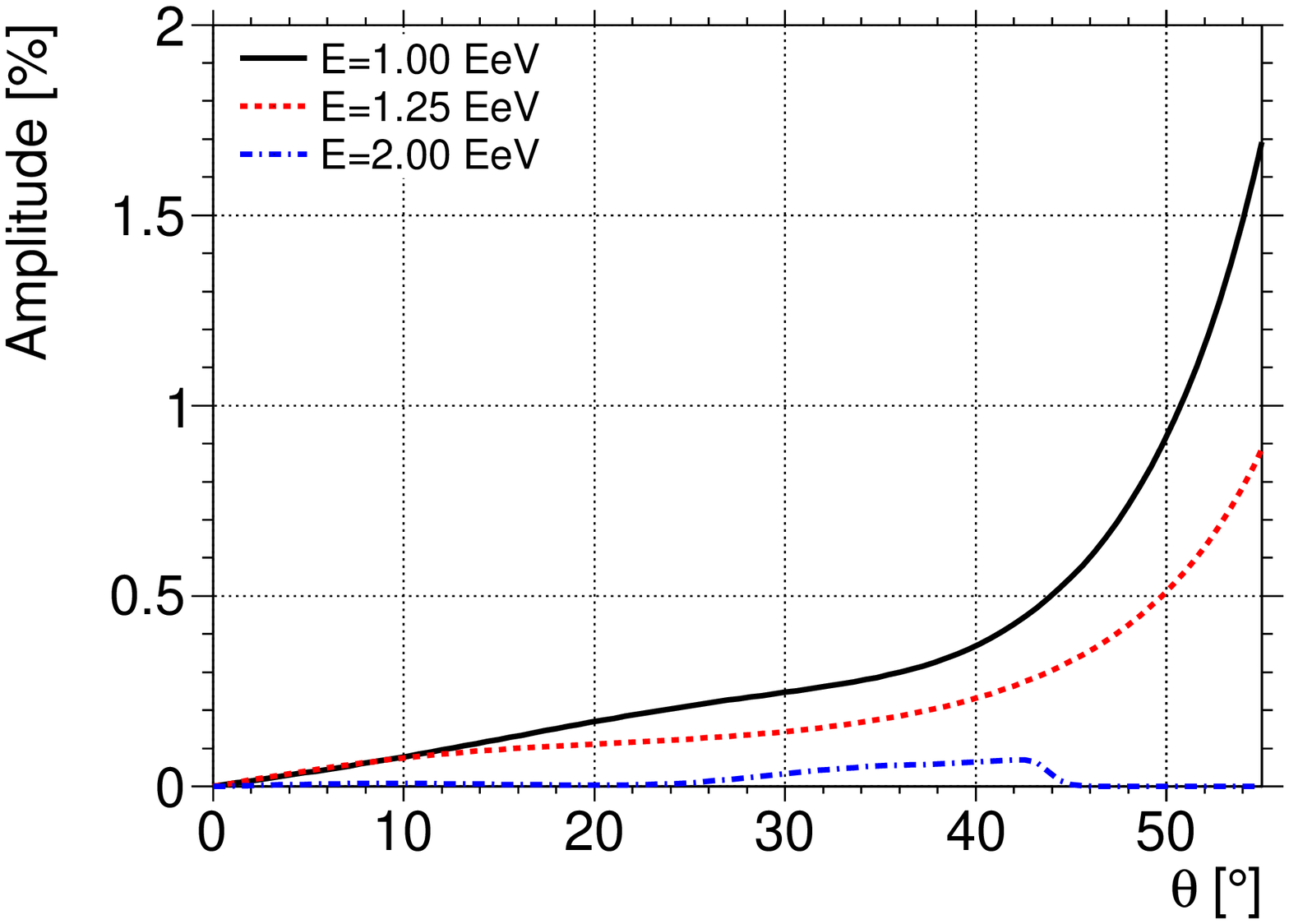}
  \caption{\small{Left~: Dependence of the detection efficiency on azimuth for $\theta=55^\circ$ and
  $E=1~$EeV, due to geomagnetic effects. Right~: Maximal contrast of the azimuthal modulation of the 
  detection efficiency induced by geomagnetic effects as a function of the zenith angle.}}
\label{fig:amp-gmf-vs-th}
\end{figure}

In addition to the effects on the energy determination presented in section~\ref{s1000-g}, geomagnetic 
effects also affect the detection efficiency for showers with energies below 3~EeV. This is because 
under any incident angles $(\theta,\varphi)$, a shower with an energy $E$ triggers the SD array with 
a probability associated with its size which is a function of azimuth because of the geomagnetic 
effects~\footnote{Here, the shorthand notation $\Delta(\theta,\varphi)$ stands for $g_1\cos^{-g_2}{(\theta)}\left[\sin^2{(\widehat{\textbf{u},\textbf{b}})}-\left<\sin^2{(\widehat{\textbf{u},\textbf{b}})}\right>_{\varphi}\right]$. The energy $E\times(1+\Delta(\theta,\varphi))^B$ is actually the 
one that would have been obtained without correcting for geomagnetic 
effects.}~: $E\times(1+\Delta(\theta,\varphi))^B$. Above 1~EeV,
this effect is in fact the main source of azimuthal dependence of the detection efficiency, so that to first 
order in $\Delta(\theta,\varphi)$, $\epsilon(\theta,\varphi,E)$ can be estimated as~:
\begin{eqnarray}
\label{eqn:eps2}
\epsilon(\theta,\varphi,E) &=& \frac{1}{\mathcal{N}}\frac{\mathrm{d}N(\sin^2{\theta},E(1+\Delta(\theta,\varphi))^B)}{\mathrm{d}\sin^2{\theta}} \nonumber \\
&\simeq&\left<\epsilon(\theta,\varphi,E)\right>_{\varphi}+\frac{BE\Delta(\theta,\varphi)}{\mathcal{N}}\frac{\partial \left<\epsilon(\theta,\varphi,E)\right>_{\varphi}}{\partial E}.
\end{eqnarray}
The correction to the detection efficiency induced by geomagnetic effects, and in particular the
azimuthal dependence, is thus straightforward to implement from the knowledge of 
$\left<\epsilon(\theta,\varphi,E)\right>_{\varphi}$. An example of such an azimuthal dependence is shown
in the left panel of Fig.~\ref{fig:amp-gmf-vs-th}, for $E=1~$EeV and $\theta=55^\circ$. The modulation 
reflects the one due to the energy determination~: the detection efficiency is lowered in the directions where
the uncorrected energies are under-estimated due to geomagnetic effects, and the efficiency is higher where 
energies are over-estimated.
The maximal contrast of such azimuthal modulations is displayed in the right panel as a function of the 
zenith angle, for three different energies. At 2~EeV, the amplitude slightly increases up to $\simeq 35^\circ$,
 staying below $\simeq 0.1\%$, and then decreases and even cancels due to the saturation of the 
detection efficiency. In contrast, when going down in energy, the relative amplitude largely 
increases with the zenith angle due to the increase of the derivative term, reaching
$\simeq 1.7\%$ for $\theta=55^\circ$ and $E=1~$EeV. 

\subsection{Tilt of the array}
\label{sub:tilt}

\begin{figure}[!t]
  \centering					 
  \includegraphics[height=8cm,width=8cm]{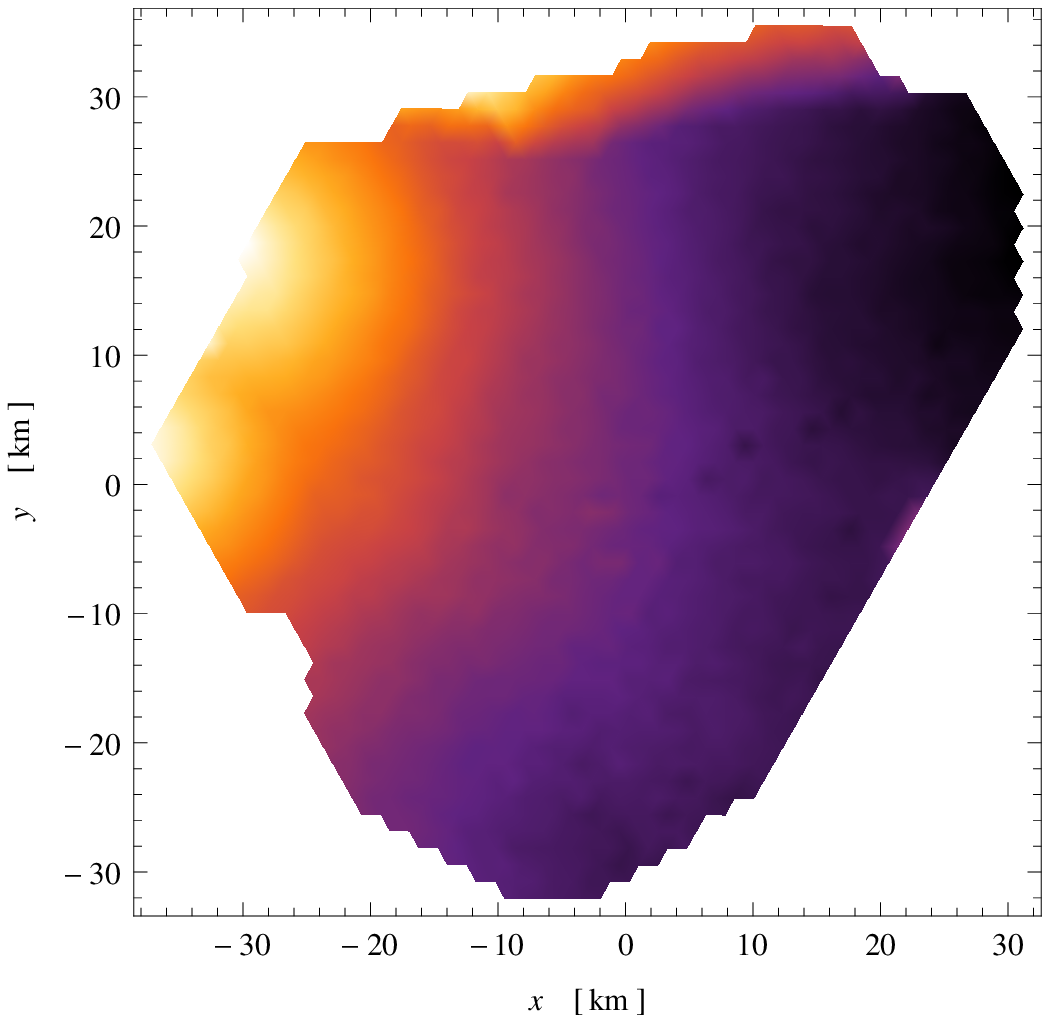}
  ~
  \raisebox{0.8cm}{
  \includegraphics[height=7.2cm]{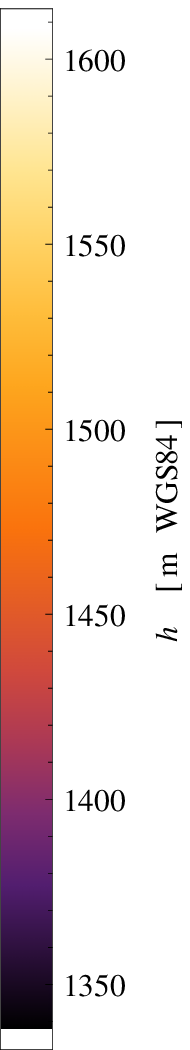}
  }
  \caption{\small{Colour-coded altitude (a.s.l.) of the water-Cherenkov detectors.}}
\label{fig:sdarray}
\end{figure}

The altitudes above sea level of the water-Cherenkov detectors are displayed in Fig.~\ref{fig:sdarray} in colour coding. The 
coordinates are in a Cartesian system whose origin is defined at the "centre" of the Observatory site. 
The Andes ridge building up in the western and north-western direction can be seen. A slightly tilted SD array 
gives rise to a small azimuthal asymmetry, and consequently slightly modifies the directional exposure 
with respect to Eqn.~\ref{eqn:omega} through small changes of the geometric directional aperture. This 
modification is twofold~: the tilt changes the geometric factor ($\cos{\theta}$) of the projected surface 
under incidence angles $(\theta,\varphi)$; and also induces a compensating effect below full efficiency by 
slightly varying the detection efficiency with the azimuth angle $\varphi$.

Denoting $\mathbf{n_\perp^{(i)}}$ the normal vector to each elemental cell, the geometric directional 
aperture per cell is not any longer simply given by $\cos{\theta}$ but now depends on both $\theta$ and $\varphi$~:
\begin{equation}
\label{eqn:tilt1}
a_{\mathrm{cell}}^{(i)}(\theta,\varphi)=1.95~ \mathbf{n}\cdot\mathbf{n_\perp^{(i)}}
\simeq 1.95~[1+\zeta^{(i)}\tan{\theta}\cos{(\varphi-\varphi_0^{(i)})}]~\cos{\theta},
\end{equation}
where $\zeta^{(i)}$ and $\varphi_0^{(i)}$ are the zenith and azimuth angles of $\mathbf{n_\perp^{(i)}}$.
It is actually this latter expression $a_{\mathrm{cell}}$ which has to be inserted into Eqn.~\ref{eqn:omega}
to calculate the directional exposure. Overall, the average tilt of the SD array is 
$\zeta^{\mathrm{eff}}\simeq 0.2^\circ$, and induces a dipolar asymmetry in azimuth with a maximum in the 
downhill direction $\varphi_0^{\mathrm{eff}}\simeq0^\circ$ and with an amplitude increasing with the zenith 
angle as $\simeq0.3\%\tan{\theta}$.

Below 3~EeV, the tilt of the array induces an additional variation of the detection efficiency with azimuth.
This is because the effective separation between detectors for a given zenith angle depends now on 
the azimuth. Since, for a given zenith angle, the SD array seen by showers coming from the uphill 
direction is denser than that for those coming from the downhill direction, the detection efficiency is 
higher in the uphill direction. Parameterising the energy dependence of $\epsilon$ as $E^3/(E^3+E_{0.5}^3)$, 
we show in Appendix~B that the change in the detection efficiency can be estimated as~:
\begin{equation}
\label{eqn:tilt2}
\Delta\epsilon_{\mathrm{tilt}}(\theta,\varphi,E) = \frac{E^3(E_{0.5}^3-{E_{0.5}^{\mathrm{tilt}}}^3(\theta,\varphi))}{(E^3+E_{0.5}^3)(E^3+{E_{0.5}^{\mathrm{tilt}}}^3(\theta,\varphi))},
\end{equation}
where $E_{0.5}^{\mathrm{tilt}}(\theta,\varphi)$ is related to $E_{0.5}$ through~:
\begin{equation}
\label{eqn:E_0.5}
E_{0.5}^{\mathrm{tilt}}(\theta,\varphi)\simeq E_{0.5}\times[1+\zeta^{\mathrm{eff}}\tan{\theta}\cos{(\varphi-\varphi_0^{\mathrm{eff}})}]^{3/2}.
\end{equation}
Around 1~EeV, this correction tends to compensate the pure geometrical effect described above, and 
even overcompensates it at lower energies.

\subsection{Spatial extension of the array}
\label{sub:extension}

This spatial extension of the SD array is such that the range of latitudes
covered by all cells reaches $\simeq 0.5^\circ$. This induces a slightly different directional
exposure between the cells located at the northern part of the array and the ones located at the
southern part. This spatial extension can be accounted for to calculate the overall directional
exposure using the cell latitudes $\ell_{\mathrm{cell}}^{(i)}$ instead of the mean site one
in the transformations from local to celestial angles in Eqn.~\ref{eqn:theta-phi}.

\subsection{Weather effects below full efficiency}
\label{sub:other}

In the same way as geomagnetic effects, weather effects can also affect the detection efficiency for 
showers with energies below 3~EeV. However, \emph{above} 1~EeV, we have shown in~\citep{AugerAPP2011} 
that as long as the analysis covers an integer number of years with almost
equal exposure in every season, the amplitude of the spurious modulation in right ascension induced 
by this effect is small enough to be neglected when performing anisotropy analyses at the present 
level of sensitivity. 

\subsection{Final estimation of the directional exposure - Examples at some energies}
\label{sub:examples}

\begin{figure}[!t]
  \centering					 
  \includegraphics[width=10cm]{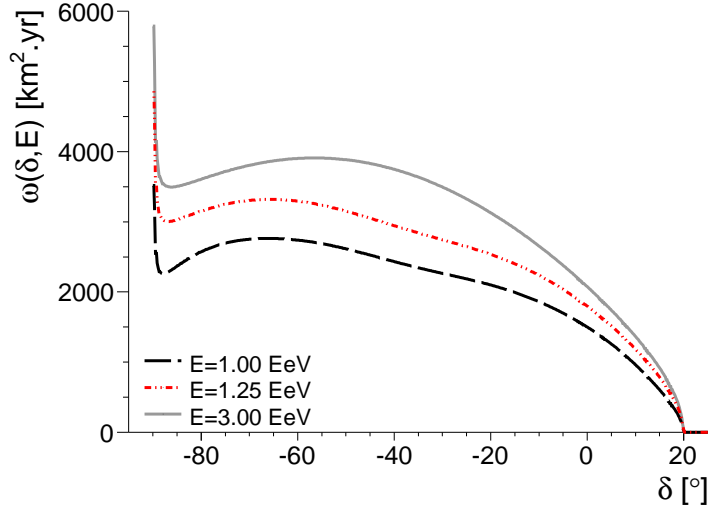}
  \caption{\small{Directional exposure $\omega(\delta,E)$ as a function of the declination $\delta$, 
  for three different energies.}}
\label{fig:expo}
\end{figure}

Accounting for all effects, the final expression to calculate the directional exposure is slightly modified
with respect to Eqn.~\ref{eqn:omega}~:
\begin{equation}
\label{eqn:omega2}
\omega(\delta,E) = \sum_{i=1}^{n_{\mathrm{cell}}}x^{(i)}\int_0^{24h}~\mathrm{d}\alpha^\prime\,a_{\mathrm{cell}}^{(i)}{(\theta,\varphi)}~\left[\epsilon(\theta,\varphi,E)+\Delta\epsilon_{\mathrm{tilt}}(\theta,\varphi,E)\right],
\end{equation}
where both $\theta$ and $\varphi$ depend on $\alpha^\prime$, $\delta$ and $\ell_{\mathrm{cell}}^{(i)}$.
The resulting dependence on declination is displayed
in Fig.~\ref{fig:expo} for three different energies. Down to 1~EeV, the detection efficiency at high zenith
angles is high enough that the equatorial south pole is visible at any time and hence constitutes the
direction of maximum of exposure. For a wide range of declinations between $\simeq -89^ \circ$ and
$\simeq -20^ \circ$, the directional exposure is $\simeq 2,500~$km$^ 2$~yr at 1~EeV, and 
$\simeq 3,500~$km$^ 2$~yr for any energy above full efficiency. Then, at higher declinations, 
it smoothly falls to zero, with no exposure above $\simeq 20^ \circ$ declination.

The average expected number of events within any solid angle and any energy range can be
recovered by integrating the directional exposure over the  solid angle 
considered and the cosmic ray energy spectrum in the corresponding energy range. Note that the rapid 
variation of the exposure close to the South pole on an angular scale of the order of the angular 
resolution has no influence on the event counting rate, due to the quasi-zero solid angle 
in that particular direction. Consequently, though the exposure around the South pole could be affected
by small changes of the detection efficiency around $\theta=55^\circ$, the results presented in next sections 
are on the other hand \textit{not} affected by the exact value of the exposure for declinations a few degrees
away from the South pole.

\section{Searches for large scale patterns}
\label{analysis}

\subsection{Estimates of spherical harmonic coefficients}

Any angular distribution over the sphere $\Phi(\mathbf{n})$ can be decomposed
in terms of a multipolar expansion~:
\begin{equation}
\label{eqn:ylm}
\Phi(\mathbf{n})=\sum_{\ell\geq0}\sum_{m=-\ell}^{\ell}~a_{\ell m}Y_{\ell m}(\mathbf{n}),
\end{equation}
where $\mathbf{n}$ denotes a unit vector taken in equatorial coordinates. The
customary recipe to extract each multipolar coefficient makes use of the completeness 
relation of spherical harmonics~:
\begin{equation}
\label{eqn:alm}
a_{\ell m}=\int_{4\pi} \mathrm{d}\Omega~\Phi(\mathbf{n})Y_{\ell m}(\mathbf{n}),
\end{equation}
where the integration is over the entire sphere of directions $\mathbf{n}$.
Any anisotropy fingerprint is encoded in the $a_{\ell m}$ spherical harmonic coefficients. 
Variations on an angular scale of $\Theta$ radians contribute amplitude in the $\ell\simeq1/\Theta$ 
modes.

However, in case of partial sky coverage, the solid angle in the sky where the exposure is zero makes it 
impossible to estimate the multipolar coefficients $a_{\ell m}$ in this way. This is because the unseen 
solid angle prevents one from making use of the completeness relation of the spherical 
harmonics~\citep{Sommers2001}. Since the observed arrival direction distribution is in this case
the \textit{combination} of the angular distribution $\Phi(\mathbf{n})$ and of the directional exposure 
function $\omega(\mathbf{n})$, the integration performed in Eqn.~\ref{eqn:alm} does not allow any 
longer the extraction of the multipolar coefficients of $\Phi(\mathbf{n})$, but only the ones of 
$\omega(\mathbf{n})~\Phi(\mathbf{n})$~\citep{alm}~\footnote{To cope with the 
unseen solid angle, another approach makes use of orthogonal functions of increasing multipolarity, 
tailored to the exposure $\omega$ itself~\citep{alm}. This method would yield similar accuracies.}:
\begin{eqnarray}
\label{eqn:blm}
b_{\ell m}&=&\int_{\Delta\Omega} \mathrm{d}\Omega~\omega(\mathbf{n})\Phi(\mathbf{n})Y_{\ell m}(\mathbf{n}) \nonumber \\
&=&\sum_{\ell^\prime\geq0}\sum_{m^\prime=-\ell^\prime}^{\ell^\prime} a_{\ell^\prime m^\prime}\int_{\Delta\Omega} \mathrm{d}\Omega~\omega(\mathbf{n})Y_{\ell^\prime m^\prime}(\mathbf{n})Y_{\ell m}(\mathbf{n}).
\end{eqnarray}
Formally, the $a_{\ell m}$ coefficients appear related to the $b_{\ell m}$ ones 
through a convolution such that $b_{\ell m}=\sum_{\ell^\prime\geq0}\sum_{m^\prime=-\ell^\prime}^{\ell^\prime}[K]_{\ell m}^{\ell^\prime m^\prime}~a_{\ell^\prime m^\prime}$. The matrix $K$, which imprints the interferences between 
modes induced by the non-uniform and partial coverage of the sky, is entirely determined by the 
directional exposure. The relationship established in Eqn.~\ref{eqn:blm} is valid for \textit{any} 
exposure function $\omega(\mathbf{n})$. 

Meanwhile, the observed arrival direction distribution, $\overline{\mathrm{d}N}(\mathbf{n})/\mathrm{d}\Omega$, 
provides a direct estimation of the $b_{\ell m}$ coefficients 
through (hereafter, we use an over-line to indicate the \emph{estimator} of any quantity)~:
\begin{equation}
\label{eqn:blm-est-1}
\overline{b}_{\ell m}=\int_{\Delta\Omega} \mathrm{d}\Omega~\frac{\overline{\mathrm{d}N}(\mathbf{n})}{\mathrm{d}\Omega}~Y_{\ell m}(\mathbf{n}),
\end{equation}
where the distribution $\overline{\mathrm{d}N}(\mathbf{n})/\mathrm{d}\Omega$ of any set of $N$ arrival 
directions $\{\mathbf{n}_1, ..., \mathbf{n}_N\}$ can be modelled as a sum of Dirac functions on the sphere.
Then, if the multipolar expansion of the angular distribution $\Phi(\mathbf{n})$ is \textit{bounded} to
$\ell_{\mathrm{max}}$, that is, if the $\Phi(\mathbf{n})$ has no higher moments than $\ell_{\mathrm{max}}$,
the first $b_{\ell m}$ coefficients with $\ell\leq\ell_{\mathrm{max}}$ are related to the non-vanishing $a_{\ell m}$ 
by the  square matrix $K_{\ell_{\mathrm{max}}}$ \textit{truncated} to $\ell_{\mathrm{max}}$. Inverting this 
truncated matrix allows us to recover the underlying $a_{\ell m}$ from the measured $b_{\ell m}$ 
(with $\ell\leq\ell_{\mathrm{max}}$)~:
\begin{equation}
\label{eqn:alm-est}
\overline{a}_{\ell m}=\sum_{\ell^\prime=0}^{\ell_{\mathrm{max}}}\sum_{m^\prime=-\ell^\prime}^{\ell^\prime} [K^{-1}_{\ell_{\mathrm{max}}}]_{\ell m}^{\ell^\prime m^\prime} \overline{b}_{\ell^\prime m^\prime}.
\end{equation}
In the case of small anisotropies $(|a_{\ell m}|/a_{00}\ll1)$, the resolution on each recovered 
$\overline{a}_{\ell m}$ coefficient is proportional to $\bigg([K^{-1}_{\ell_{\mathrm{max}}}]_{\ell m}^{\ell m}\bigg)^ {0.5}$~\citep{alm}~:
\begin{equation}
\label{eqn:rms-alm}
\sigma_{\ell m}=\bigg([K^{-1}_{\ell_{\mathrm{max}}}]_{\ell m}^{\ell m}~\overline{a}_{00}\bigg)^{0.5}.
\end{equation}
The dependence on $\ell_{\mathrm{max}}$ of the coefficients of $K^{-1}_{\ell_{\mathrm{max}}}$ induces an 
intrinsic indeterminacy of each recovered coefficient $\overline{a}_{\ell m}$ as $\ell_{\mathrm{max}}$ is 
increasing. This is nothing else but the mathematical translation of it being impossible to know the angular 
distribution of cosmic rays in the uncovered region of the sky. 

Henceforth, we adapt this general formalism to the search for anisotropies in Auger data in different 
energy intervals. We assume that the energy dependence of the angular distribution of cosmic
rays is smooth enough that the multipolar coefficients can be considered constant for any
energy $E$ within a narrow interval $\Delta E$. The directional exposure is hereafter considered as 
independent of the right-ascension, as defined in section~\ref{exposure}. Within an energy interval 
$\Delta E$, the expected arrival direction distribution thus reads~:
\begin{equation}
\label{eqn:dNdn}
\frac{\mathrm{d}N(\mathbf{n})}{\mathrm{d}\Omega}\propto \tilde{\omega}(\delta)~\sum_{\ell\geq0}\sum_{m=-\ell}^{\ell}~a_{\ell m}Y_{\ell m}(\mathbf{n}),
\end{equation}
where $\tilde{\omega}(\delta)$ is the effective directional exposure for the energy interval $\Delta E$. 
For convenience, this latter function is normalised such that~:
\begin{equation}
\tilde{\omega}(\delta)= \frac{\displaystyle\int_{\Delta E} \mathrm{d}E~E^{-\gamma}\omega(\delta,E)}{\displaystyle \max_\delta\bigg[\int_{\Delta E} \mathrm{d}E~E^{-\gamma}\omega(\delta,E)\bigg]},
\end{equation}
with $\gamma$ the spectral index in the considered energy range.
This dimensionless function provides, for any direction on the sky, the effective directional exposure in the 
energy range $\Delta E$ at that direction, \textit{relative} to the largest directional exposure on the sky. 
This is actually the relevant quantity which enters into Eqn.~\ref{eqn:blm} for the analyses presented below. 
Note that for a directional exposure independent of the right ascension, the coefficients 
$[K]_{\ell m}^{\ell^\prime m^\prime}$ are proportional to $\delta_m^{m^\prime}$ - \textit{i.e.} different 
values of $m$ are not mixed in the matrix. 
The observed arrival direction distribution, $\overline{\mathrm{d}N}(\mathbf{n})/\mathrm{d}\Omega$, 
is here modelled as a sum of Dirac functions on the sphere weighted by the factor 
$\Delta N_{\mathrm{cell}}^{-1}(\alpha^0_{k})$ for each event recorded at local sidereal time $\alpha^0_{k}$, 
as described in section~\ref{sub:exposure} to correct for the slightly non-uniform directional exposure in right 
ascension. In this way, the integration in Eqn.~\ref{eqn:blm} yields to~:
\begin{equation}
\label{eqn:blm-est}
\overline{b}_{\ell m}=\sum_{k=1}^{N} \frac{Y_{\ell m}(\mathbf{n}_k)}{\Delta N_{\mathrm{cell}}(\alpha^0_{k})}.
\end{equation}
The multipolar coefficients $\overline{a}_{\ell m}$ are then recovered by means of Eqn.~\ref{eqn:alm-est}.
Given the exposure functions described in section~\ref{exposure}, the resolution on each recovered coefficient,
encoded in Eqn.~\ref{eqn:rms-alm}, is degraded by a factor larger than 2 each time $\ell_{\mathrm{max}}$ is 
incremented by 1. This prevents the recovery of each coefficient with good accuracy as soon as 
$\ell_{\mathrm{max}}\geq3$, since, for $\ell_{\mathrm{max}}=3$ for instance, our current statistics 
would only allow us to probe dipole amplitudes at the 10\% level. Consequently, in the following, we restrict 
ourselves to reporting results on individual coefficients obtained when assuming a dipolar distribution 
$(\ell_{\mathrm{max}}=1)$ and a quadrupolar distribution $(\ell_{\mathrm{max}}=2)$. Meanwhile, due to
the interferences between modes induced by the non-uniform and partial sky coverage, it is important to 
stress again that each multipolar coefficient recovered under the assumption of a particular bound 
$\ell_{\mathrm{max}}$ might be biased if the underlying angular distribution of cosmic rays is not
bounded to $\ell_{\mathrm{max}}$. Given the directional exposure functions considered in this study, this
effect can be important only if the angular distribution has in fact significant moments of order 
$\ell_{\mathrm{max}}+1$.

\subsection{Searches for dipolar patterns}
\label{dipole}

As outlined in the introduction, a measurable dipole is regarded as a likely possibility 
in many scenarios for the origin of cosmic rays at EeV energies.  
Assuming that the angular distribution of cosmic rays is modulated by a \emph{pure}  
dipole, the intensity $\Phi(\mathbf{n})$ can be parameterised in any direction $\mathbf{n}$ as~:
\begin{equation}
\label{eqn:phi-dip}
\Phi(\mathbf{n})=\frac{\Phi_0}{4\pi}~\bigg(1+r~\mathbf{d}\cdot\mathbf{n} \bigg),
\end{equation}
where $\mathbf{d}$ denotes the dipole unit vector. 
The dipole pattern is here fully characterised by a declination $\delta_d$, a right ascension $\alpha_d$,
and an \emph{amplitude} $r$ corresponding to the maximal anisotropy contrast~:
\begin{equation}
r=\frac{\Phi_{\mathrm{max}}-\Phi_{\mathrm{min}}}{\Phi_{\mathrm{max}}+\Phi_{\mathrm{min}}}.
\end{equation}
The estimation of these three coefficients is straightforward from the 
estimated spherical harmonic coefficients $\overline{a}_{1m}$~: 
$\overline{r}=[3(\overline{a}_{10}^2+\overline{a}_{11}^2+\overline{a}_{1-1}^2)]^{0.5}/\overline{a}_{00}$, 
$\overline{\delta}=\arcsin{(\sqrt{3}\overline{a}_{10}/\overline{a}_{00}\overline{r})}$, and
$\overline{\alpha}=\arctan{(\overline{a}_{1-1}/\overline{a}_{11})}$.
Uncertainties on $\overline{r}$, $\overline{\delta}$ and $\overline{\alpha}$ are obtained 
from the propagation of uncertainties 
on each recovered $\overline{a}_{1m}$ coefficient (\textit{cf} Eqn.~\ref{eqn:rms-alm}).
Under an underlying isotropic distribution, and for an axisymmetric directional exposure
around the axis defined by the North and South equatorial poles, the probability density 
function of $\overline{r}$ is given by~\citep{AugerJCAP2011}~:
\begin{equation}
p_R(\overline{r})=\frac{\overline{r}}{\sigma\sqrt{\sigma_z^2-\sigma^2}}~\mathrm{erfi}\bigg(\frac{\sqrt{\sigma_z^2-\sigma^2}}{\sigma\sigma_z}\frac{\overline{r}}{\sqrt{2}}\bigg)~\exp{\bigg(-\frac{\overline{r}^2}{2\sigma^2}\bigg)},
\end{equation}
where $\mathrm{erfi}(z)=\mathrm{erf}(iz)/i$, $\sigma=\sqrt{3}\sigma_{11}/\overline{a}_{00}$, 
and $\sigma_z=\sqrt{3}\sigma_{10}/\overline{a}_{00}$. The probability $P_R(>\overline{r})$ that an amplitude 
equal or larger than $\overline{r}$ arises from a statistical fluctuation of an isotropic distribution
is then obtained by integrating $p_R$ above $\overline{r}$~:
\begin{equation}
P_R(>\overline{r})=\mathrm{erfc}\bigg(\frac{\overline{r}}{\sqrt{2}\sigma_z}\bigg)+\mathrm{erfi}\bigg(\frac{\sqrt{\sigma_z^2-\sigma^2}}{\sigma\sigma_z}\frac{\overline{r}}{\sqrt{2}}\bigg)~\exp{\bigg(-\frac{\overline{r}^2}{2\sigma^2}\bigg)}.
\end{equation}

\begin{figure}[!t]
  \centering					 
  \includegraphics[width=8.cm]{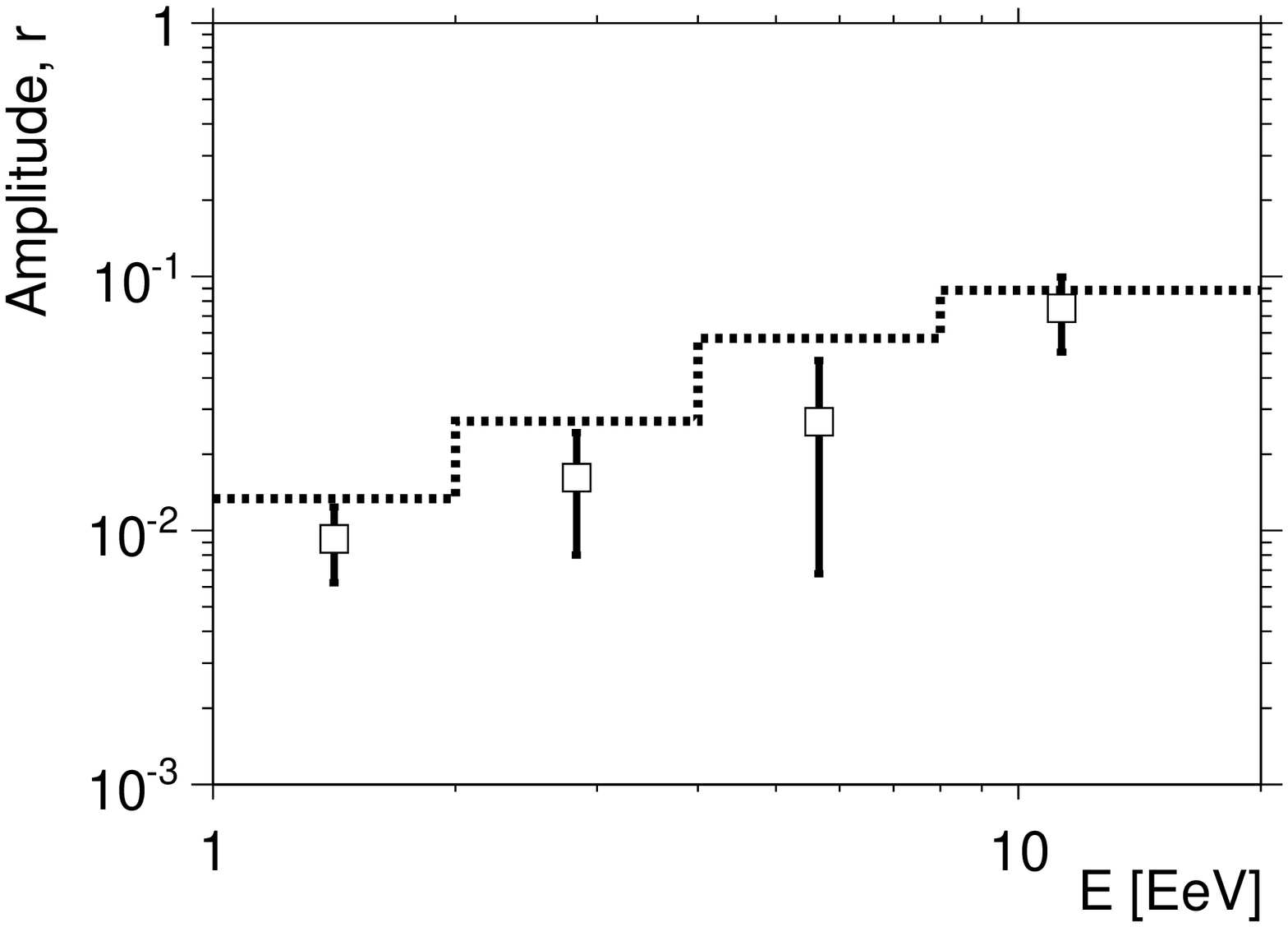}
  \includegraphics[width=7.cm]{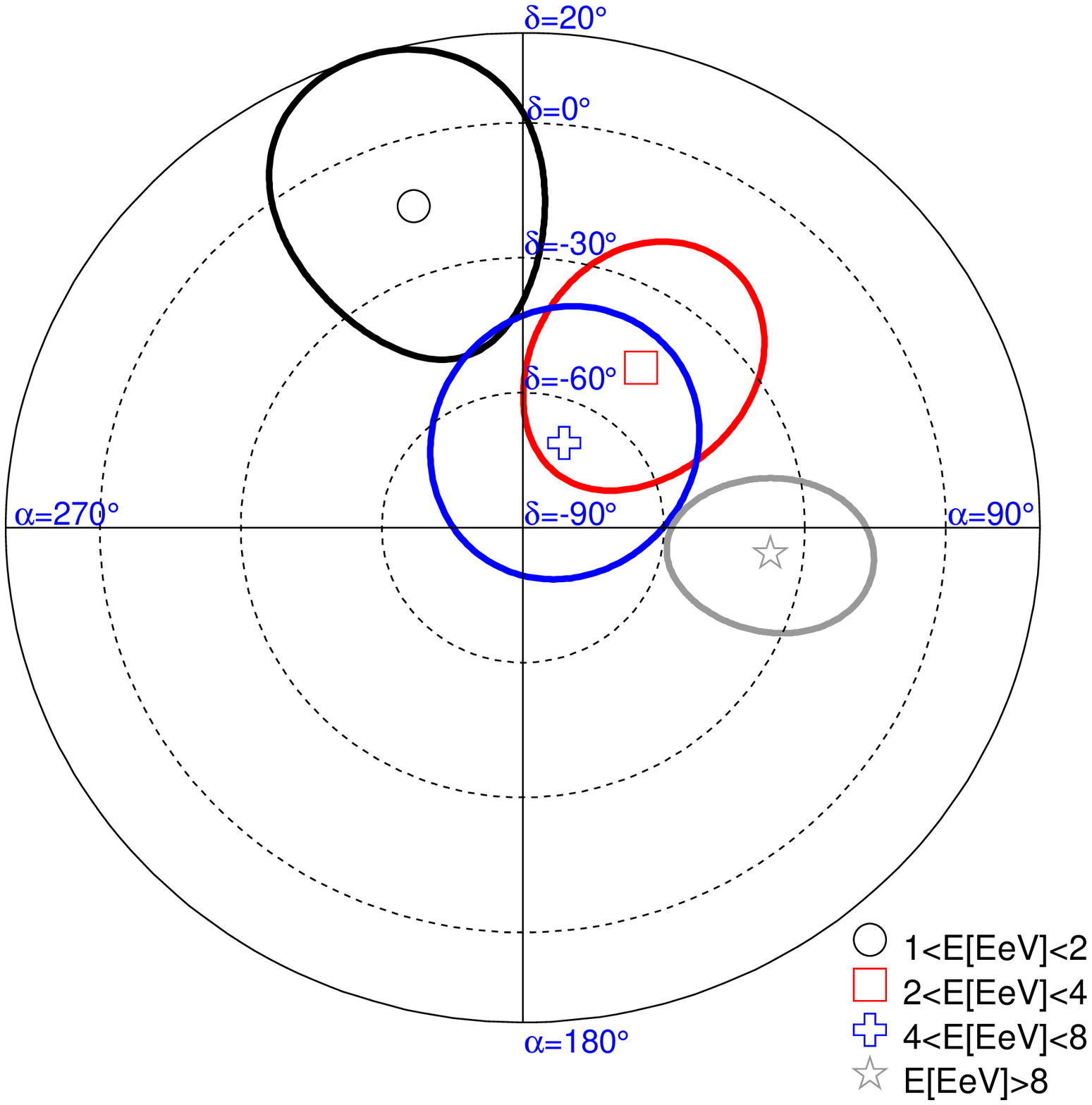}
  \caption{\small{Left~: Reconstructed amplitude of the dipole as a function of energy. The dotted line stands 
  for the 99\% $C.L.$ upper bounds on the amplitudes that would result from fluctuations of an isotropic distribution. 
  Right~: Reconstructed declination and right-ascension  of the dipole with corresponding uncertainties, as a 
  function of energy, in azimuthal projection.}}
\label{fig:dip}
\end{figure}

\begin{figure}[!t]
  \centering					 
  \includegraphics[width=7.5cm]{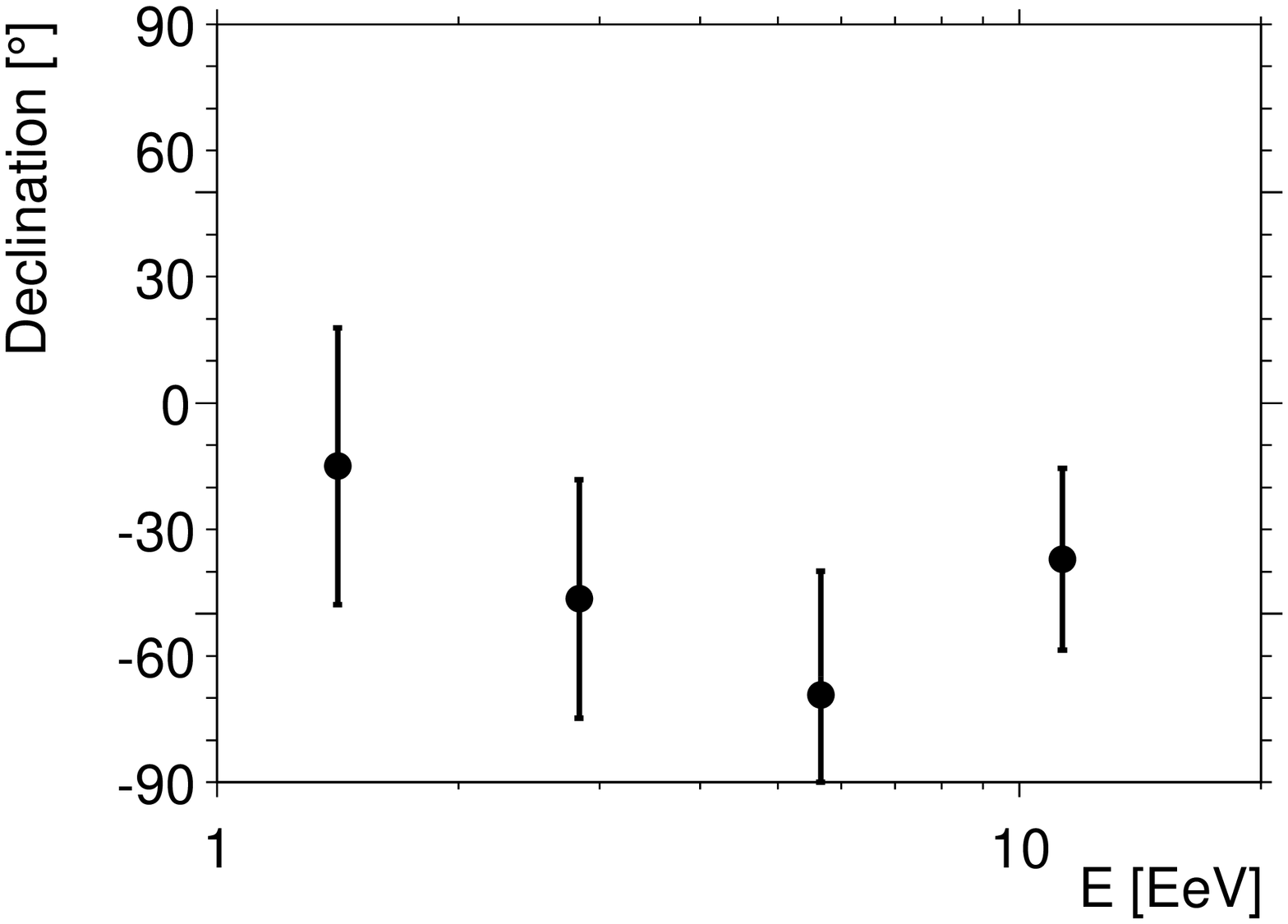}
  \includegraphics[width=7.5cm]{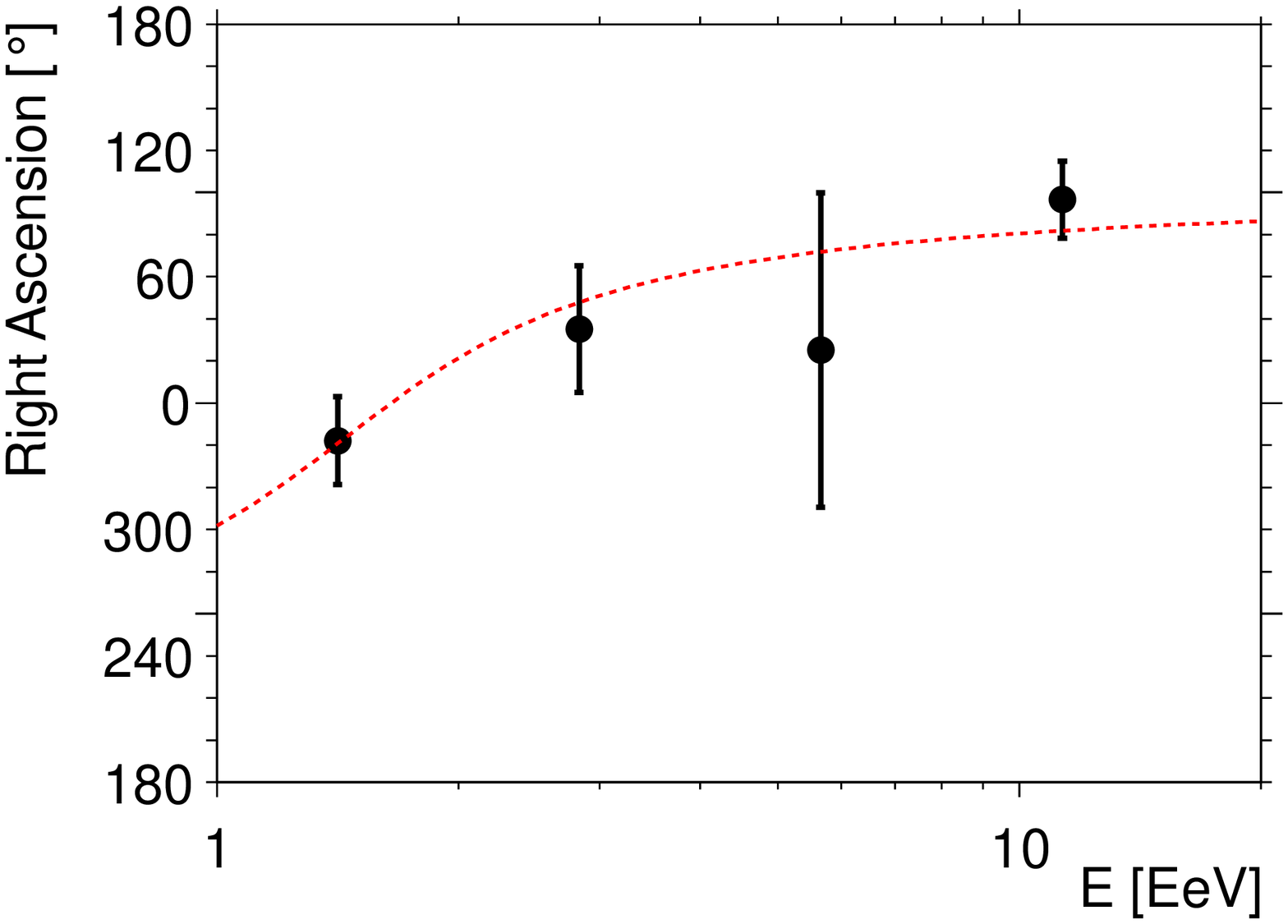}
  \caption{\small{Reconstructed declination (left) and right ascension (right) of the dipole as a function of energy.
  The smooth fit to the data of~\citep{AugerAPP2011} is shown as the dashed line in the right panel~: a consistent
  smooth behaviour is observed using the analysis presented here and applied to a data set containing two additional 
  years of data.}}
\label{fig:directiondip}
\end{figure}

The reconstructed amplitudes $\overline{r}(E)$ and corresponding directions are shown in 
Fig.~\ref{fig:dip} with the associated uncertainties, as a function of the energy. The directions
are drawn in azimuthal projection, with the equatorial South pole located at the centre and 
the right-ascension going from 0 to 360$^\circ$ clockwise. In the left panel, the 99\% $C.L.$ 
upper bounds on the amplitudes that would result from fluctuations of an isotropic distribution 
are indicated by the dotted line (\textit{i.e.} the amplitudes $\overline{r}_{99}(E)$ such 
that $P_R(>\overline{r}_{99}(E))=0.01$). One can see that within the statistical uncertainties, 
there is no strong evidence of any significant signal.

\begin{figure}[!t]
  \centering					 
  \includegraphics[width=7.5cm]{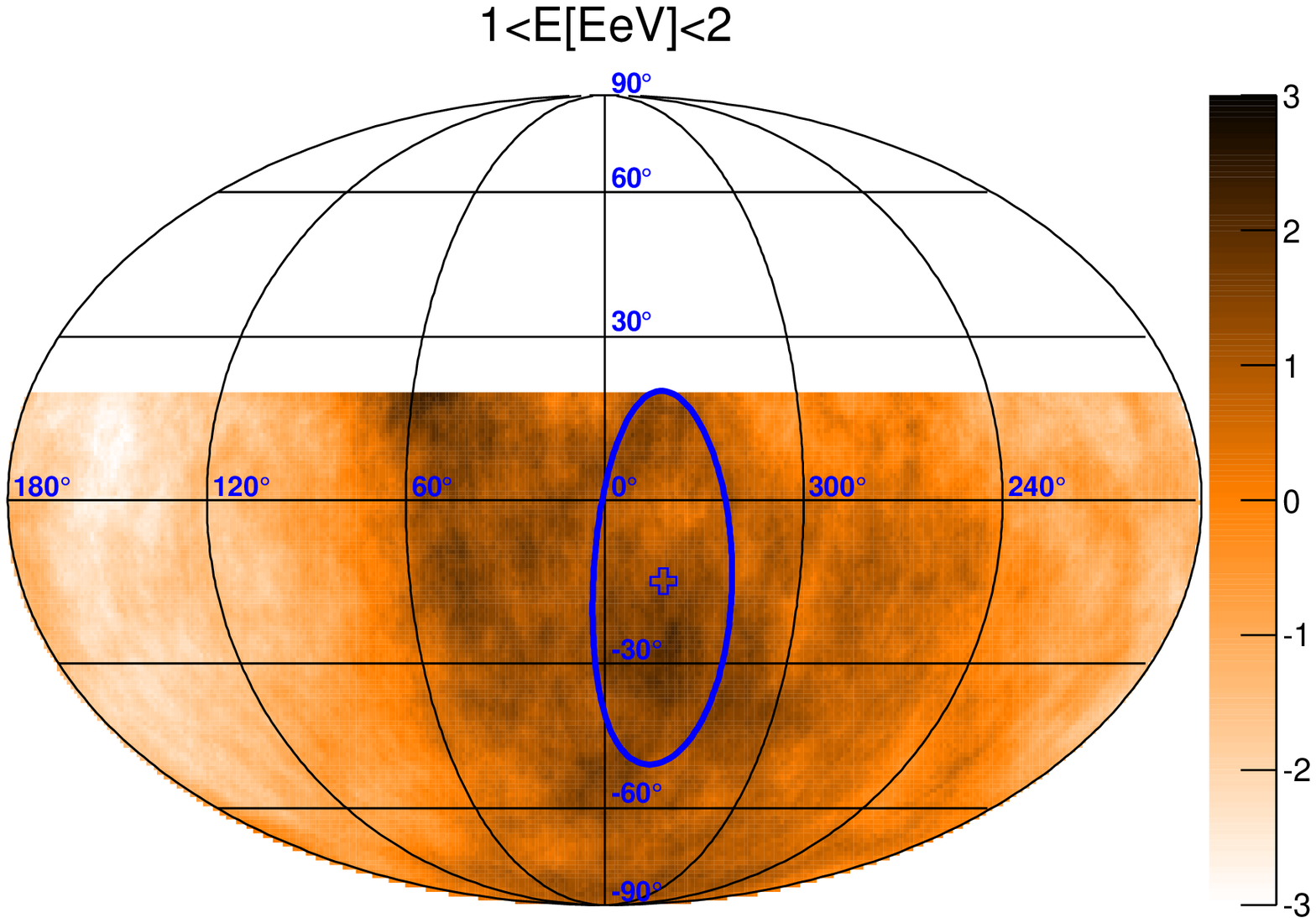}
  \includegraphics[width=7.5cm]{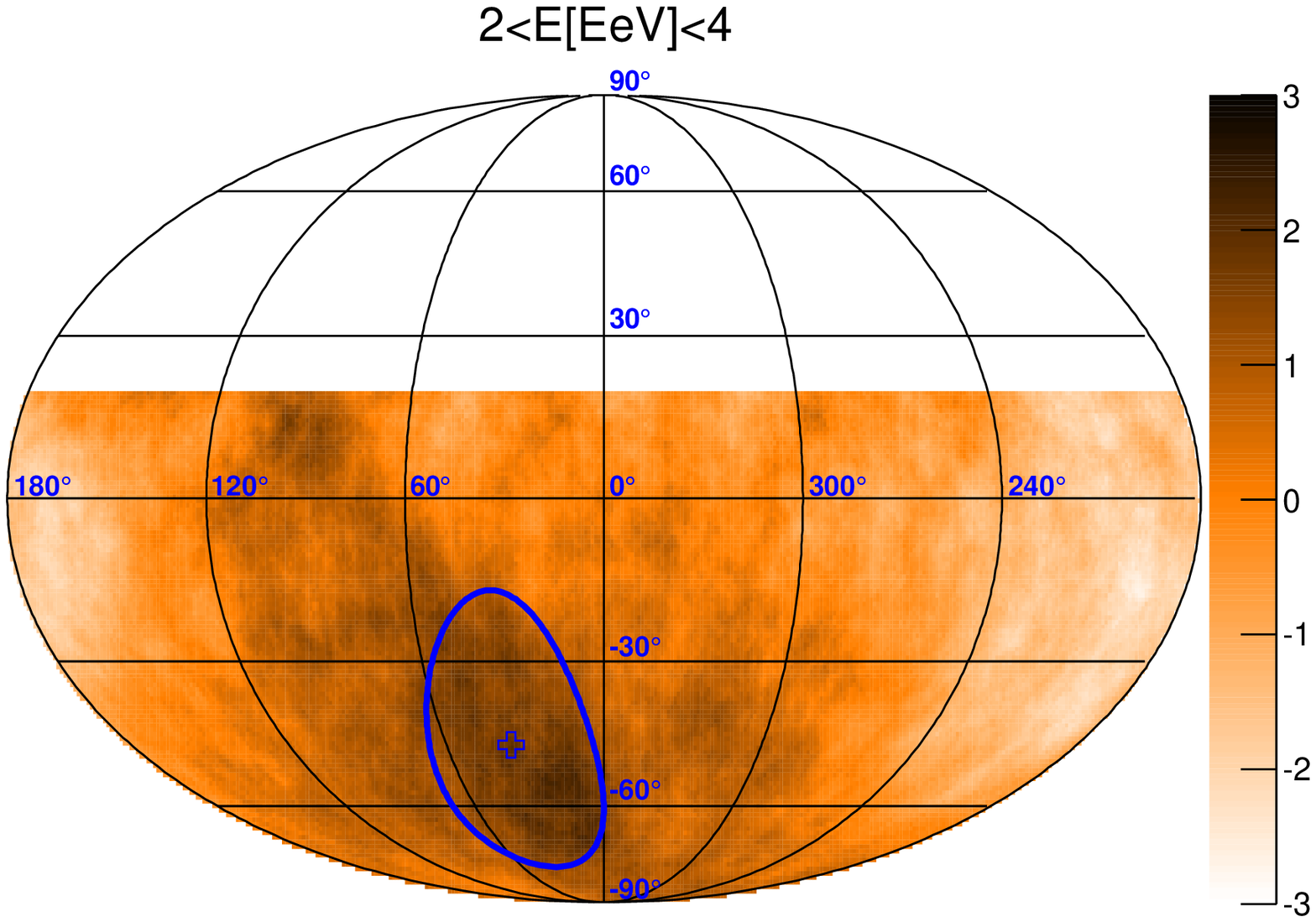}
  \includegraphics[width=7.5cm]{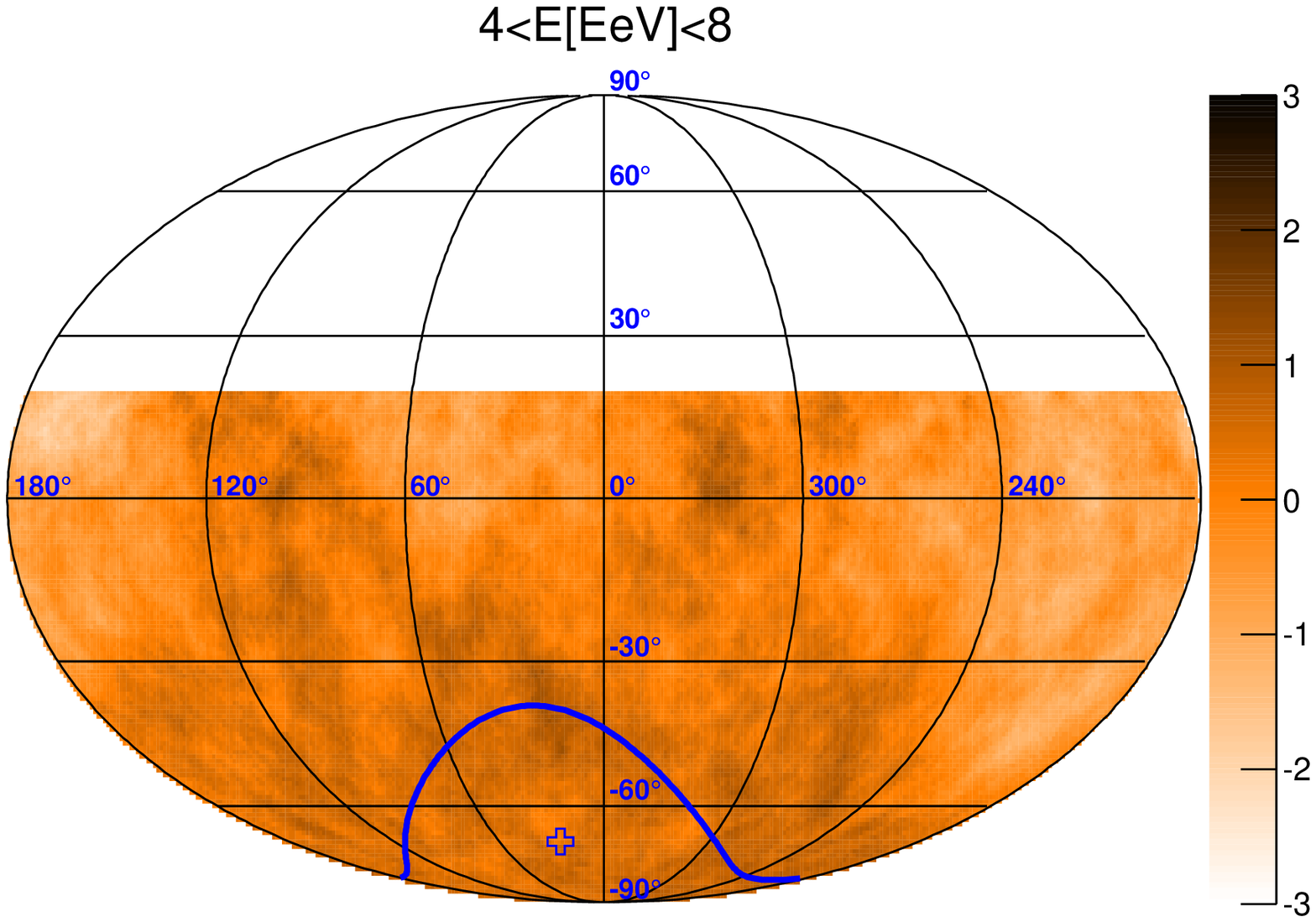}
  \includegraphics[width=7.5cm]{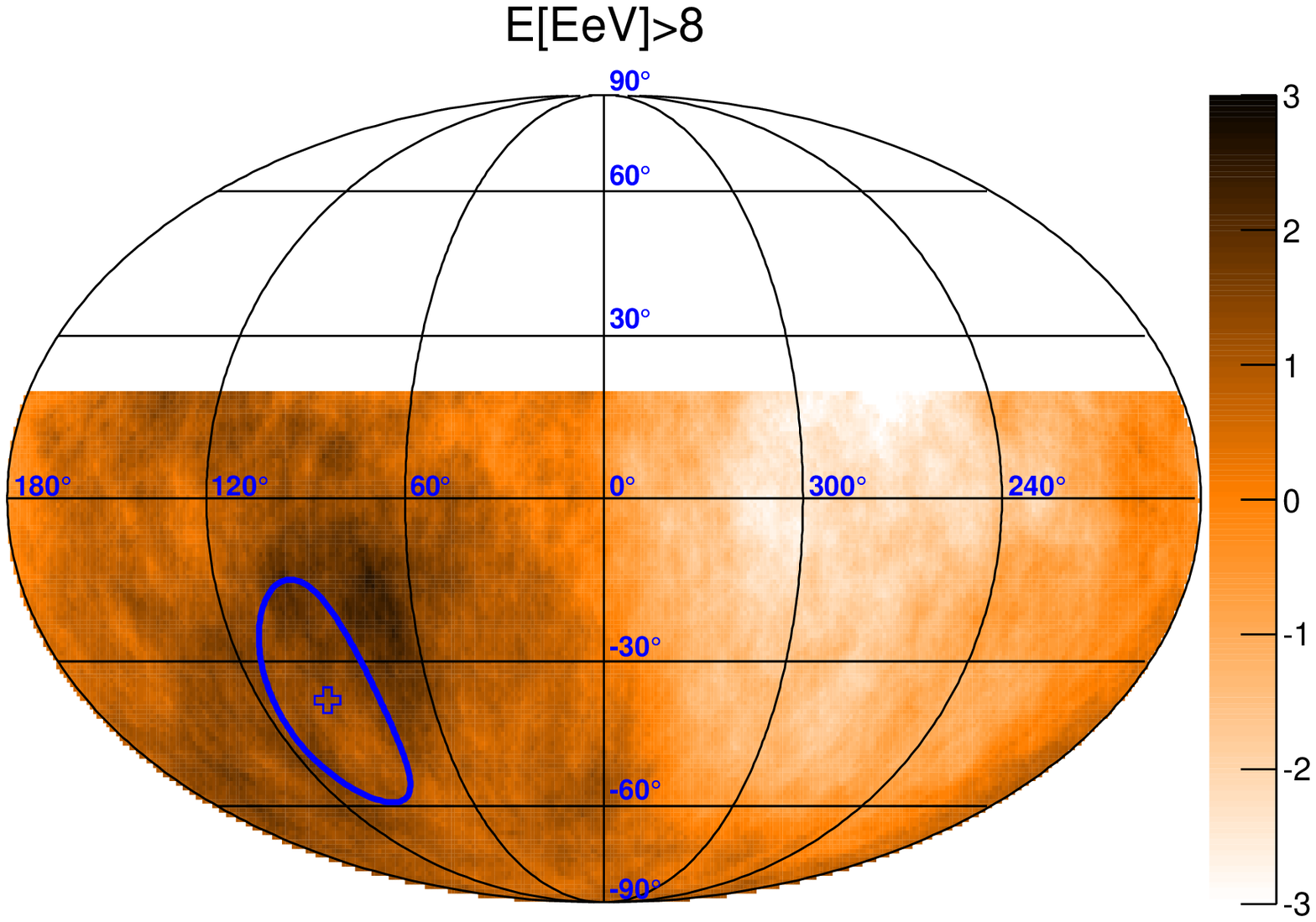}
  \caption{\small{Significance sky maps in four independent energy bins. The maps are smoothed using
  an angular window with radius $\Theta=1~$radian, to exhibit any dipolar-like structures. The directions
  of the reconstructed dipoles are shown with the associated uncertainties. The galactic plane and galactic
  center are also depicted as the dotted line and the star.}}
\label{fig:maps}
\end{figure}

The reconstructed declinations $\overline{\delta}$ and right ascensions $\overline{\alpha}$ 
are shown separately in Fig~\ref{fig:directiondip}. Both quantities are expected to be randomly 
distributed in case of independent samples whose parent distribution is isotropic.  In our previous report 
on first harmonic analysis in right ascension~\citep{AugerAPP2011}, we pointed out the intriguing smooth 
alignment of the phases in right ascension as a function of the energy, and noted that such a consistency 
of phases in adjacent energy intervals is expected to manifest with smaller number of events than those
required for the detection of 
amplitudes standing-out significantly above the background noise in case of a real underlying anisotropy. This 
motivated us to design a \textit{prescription} aimed at establishing at 99\% $C.L.$ whether this consistency 
in phases is real, using the exact same analysis as the one reported in Ref.~\citep{AugerAPP2011}. The 
prescribed test will end once the total exposure since 25 June 2011 is 21,000~km$^2$~yr~sr.
The smooth fit to the data of Ref.~\citep{AugerAPP2011} is shown as a dashed line in the right panel of 
Fig~\ref{fig:directiondip}, restricted to the energy range considered here. Though the phase between 
4 and 8~EeV is poorly determined due to the corresponding direction in declination pointing close to the 
equatorial south pole, it is noteworthy that a consistent smooth behaviour is observed using the analysis 
presented here and applied to a data set containing two additional years of data. It is also interesting 
to see in the left panel that all reconstructed declinations are in the equatorial southern hemisphere. 

\begin{figure}[!t]
  \centering					 
  \includegraphics[width=7.5cm]{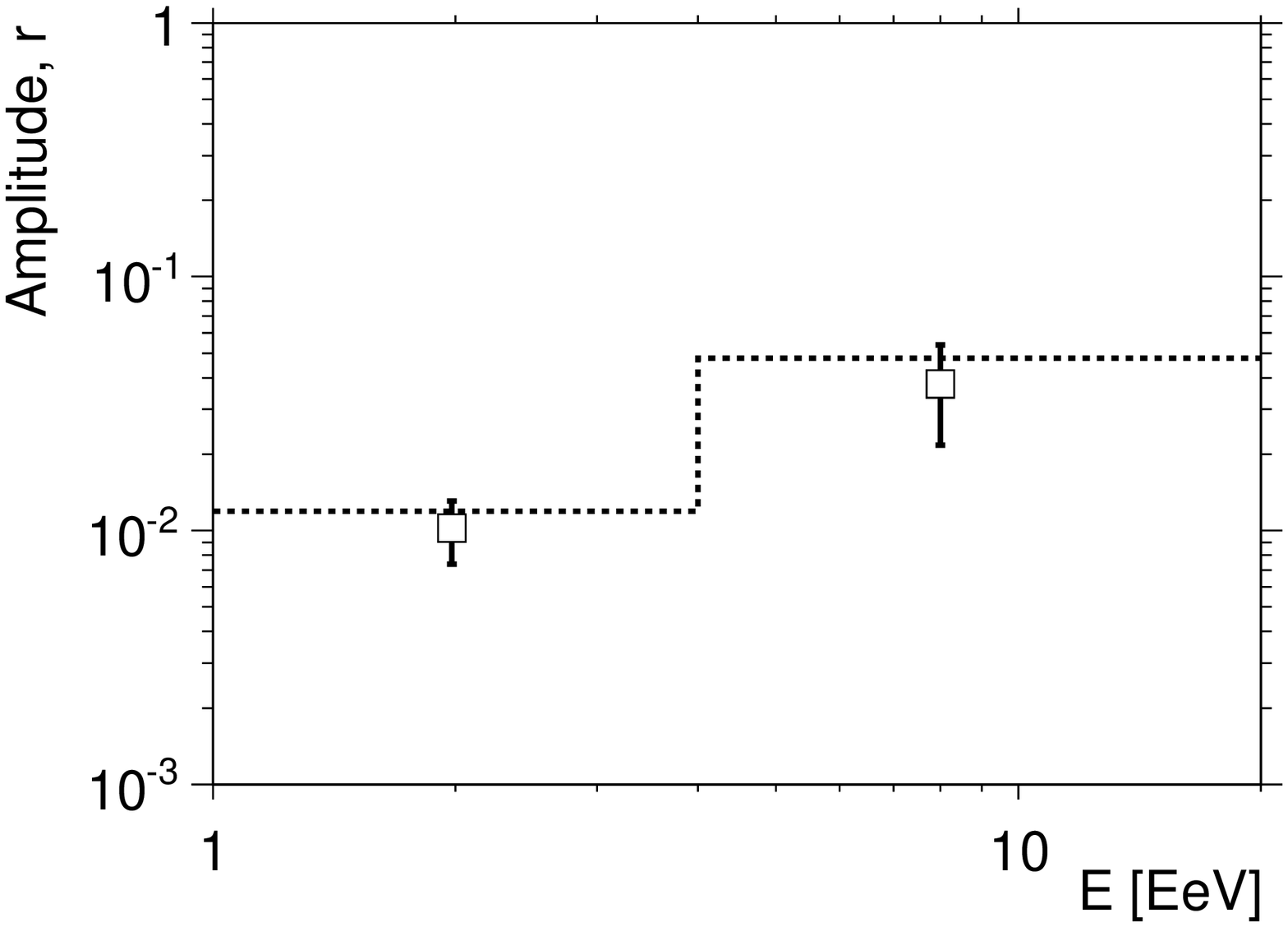}
  \includegraphics[width=7.5cm]{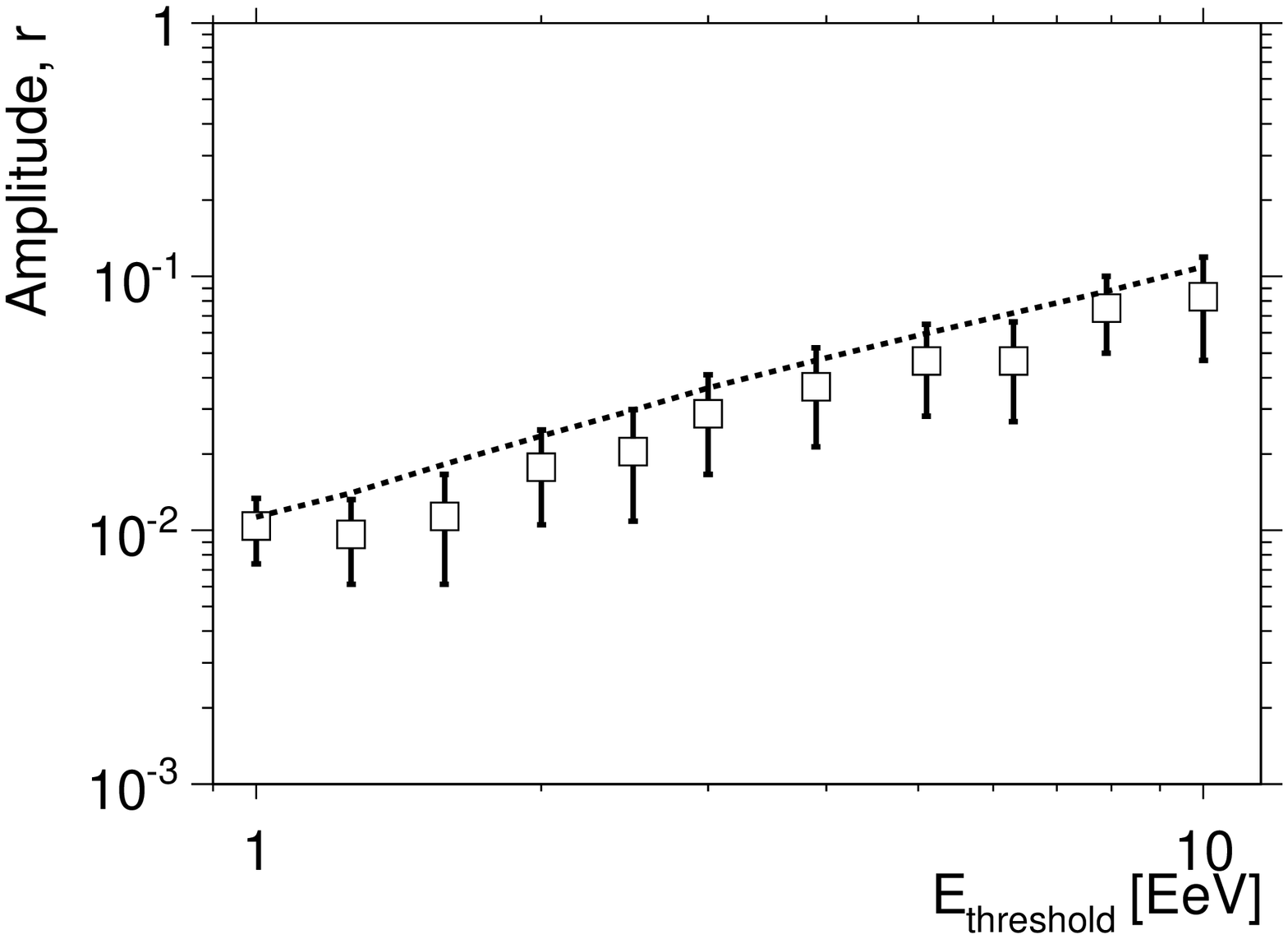}
  \caption{\small{Left~: Amplitude of the dipole for two energy intervals~: $1<E/[\mathrm{EeV}]<4$ and $E>4~$EeV.
  Right~: Amplitude of the dipole as a function of energy thresholds. The dotted lines stand for the 99\% $C.L.$ 
  upper bounds on the amplitudes that could result from fluctuations of an isotropic distribution.}}
\label{fig:rdip2}
\end{figure}

For completeness, significance sky maps are displayed in Fig.~\ref{fig:maps} in equatorial 
coordinates and using a Mollweide projection, for the four energy ranges. The galactic plane and 
galactic center are also depicted as the dotted line and the star. Significances are 
calculated using the Li and Ma estimator~\citep{LiMa1983}. This widely used estimator 
of significance, $S$, properly accounts for the fluctuations of the background and of an
eventual signal in any angular region searched~\footnote{The parameter $\alpha_{LM}$ in 
the expression of the Li \& Ma significance, expressing the expected ratio of the count numbers 
between the angular region searched (the \textit{on-region}) and any background region if there 
is no signal in the on-region, is here taken as the ratio between the expected number of events
in the on-region and the total number of events in the energy range considered.}. 
If no signal is present, the variable $S$
is nearly normally distributed even for small count numbers, so that positive values of $S$
can be interpreted as the number of standard deviations of any excess in the sky. As
well, for negative values of $S$, $-S$ can be interpreted as the number of standard 
deviations of any deficit in the sky. The maps show the overdensities obtained  
in circular windows of radius $\Theta=1~$radian, to better exhibit possible dipolar-like structures. 
The directions of the reconstructed dipoles are also shown, with their associated uncertainties
(thick circles).

Finally, since some consistency is observed both in declination and right ascension as a function of energy, 
the use of larger energy intervals and/or energy thresholds may help to pick up a significant signal above
the background level. The amplitudes of the dipole are shown in Fig.~\ref{fig:rdip2} for two energy 
intervals ($1<E/[\mathrm{EeV}]<4$ and $E>4~$EeV) and as a function of energy thresholds. 
This does not provide any further evidence for significant anisotropies.

\subsection{Searches for quadrupolar patterns}
\label{quadrupole}

Any excesses along a plane would show up as a prominent quadrupole moment. 
Such excesses are plausible for instance at EeV energies in case of an
emission of light EeV-cosmic rays from sources preferentially located in the galactic
disk, or at higher energies from sources preferentially located in the super-galactic
plane. Consequently, a measurable quadrupole may be regarded as an interesting outcome
of an anisotropy search at ultra high energies.

\begin{figure}[!t]
  \centering					 
  \includegraphics[width=7.5cm]{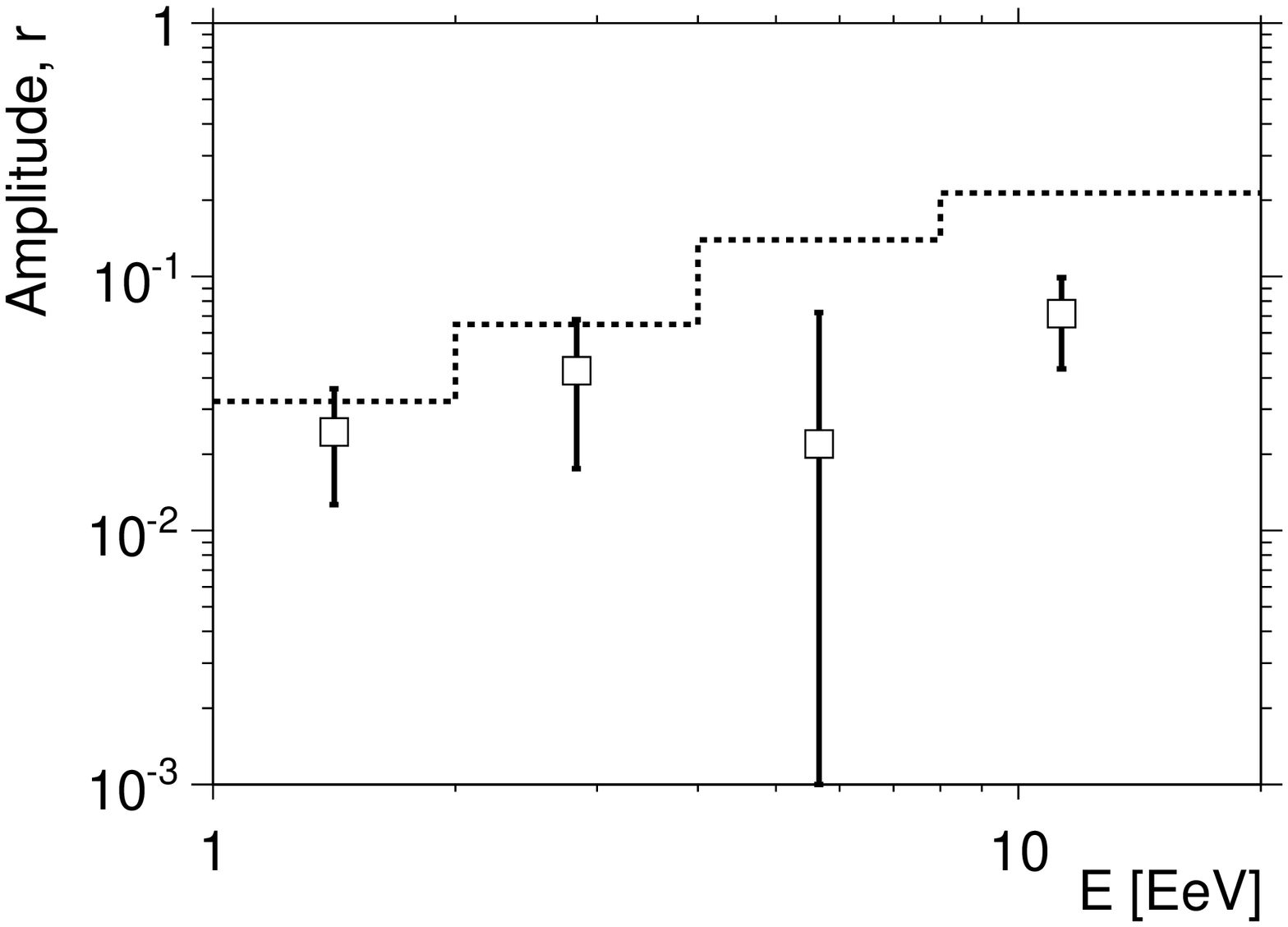}
  \includegraphics[width=7.5cm]{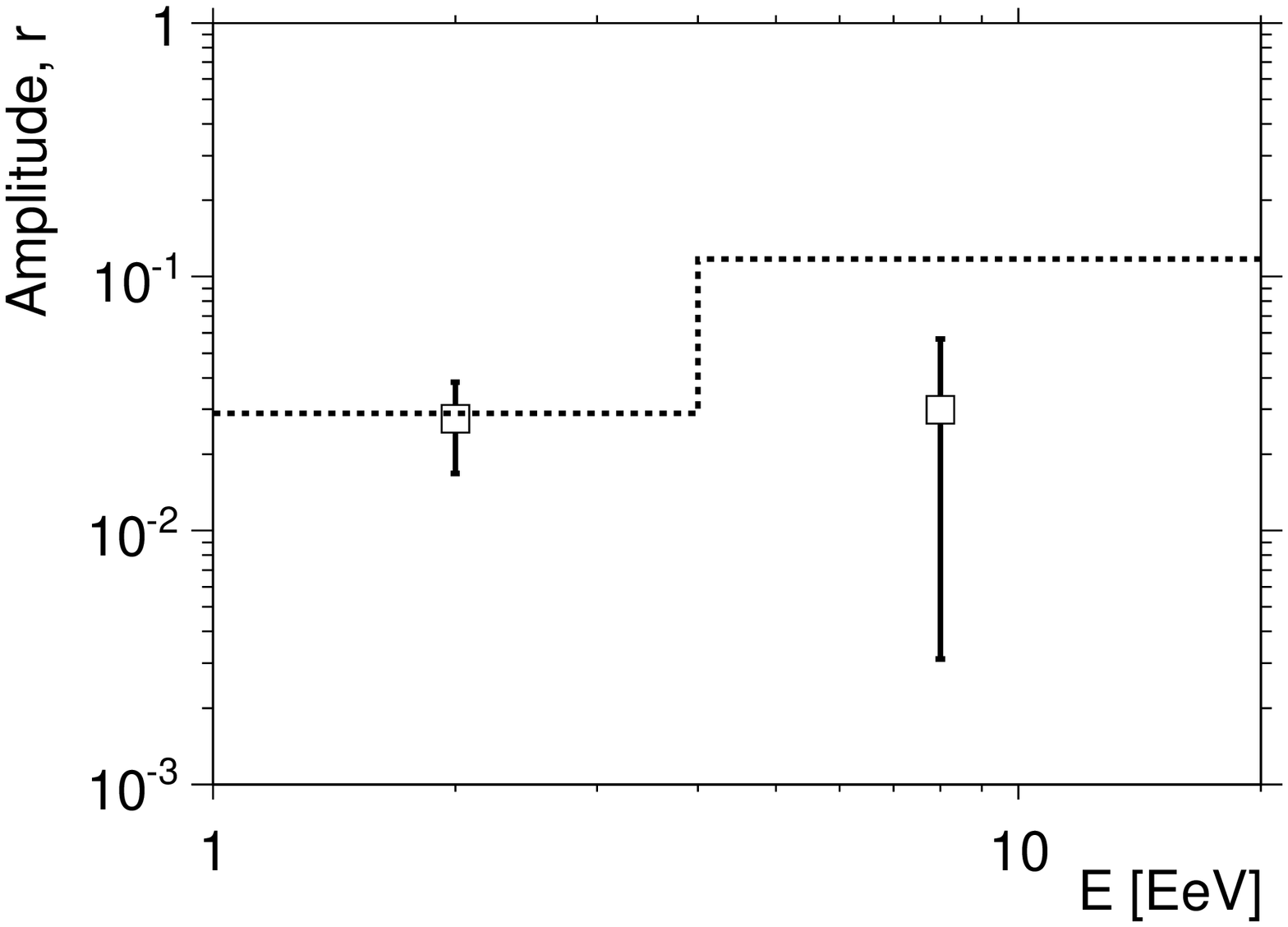}
  \includegraphics[width=7.5cm]{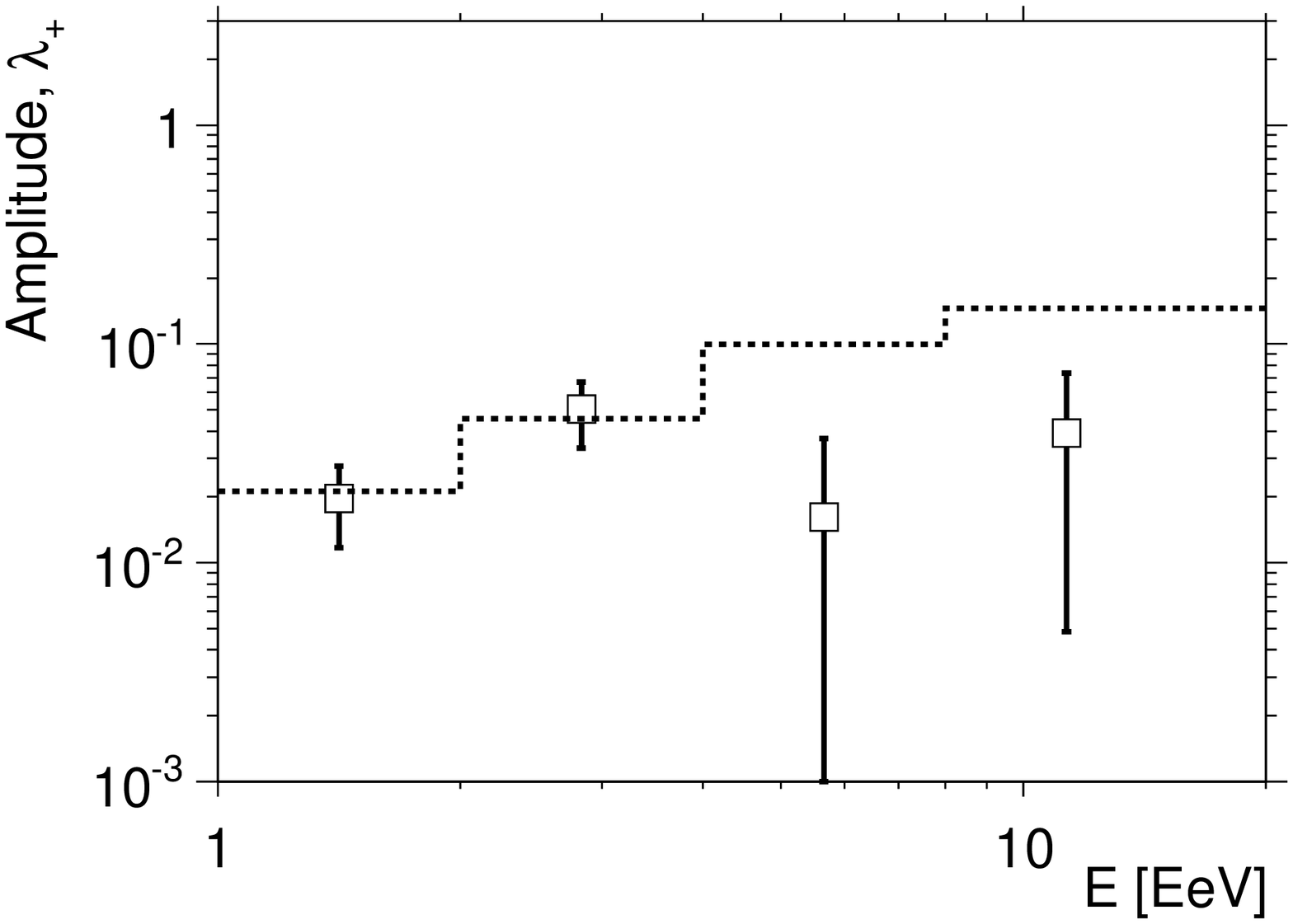}
  \includegraphics[width=7.5cm]{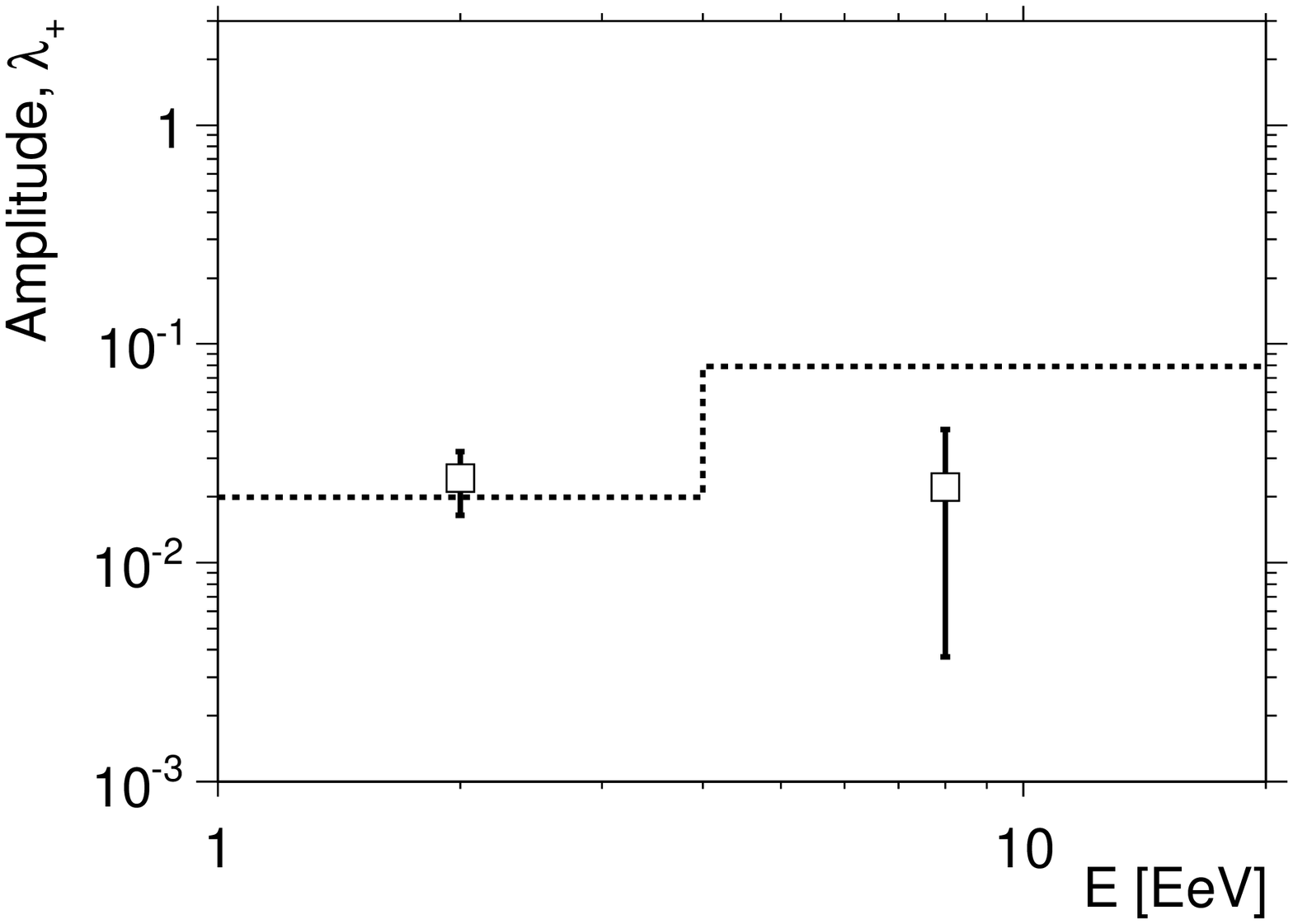}
  \includegraphics[width=7.5cm]{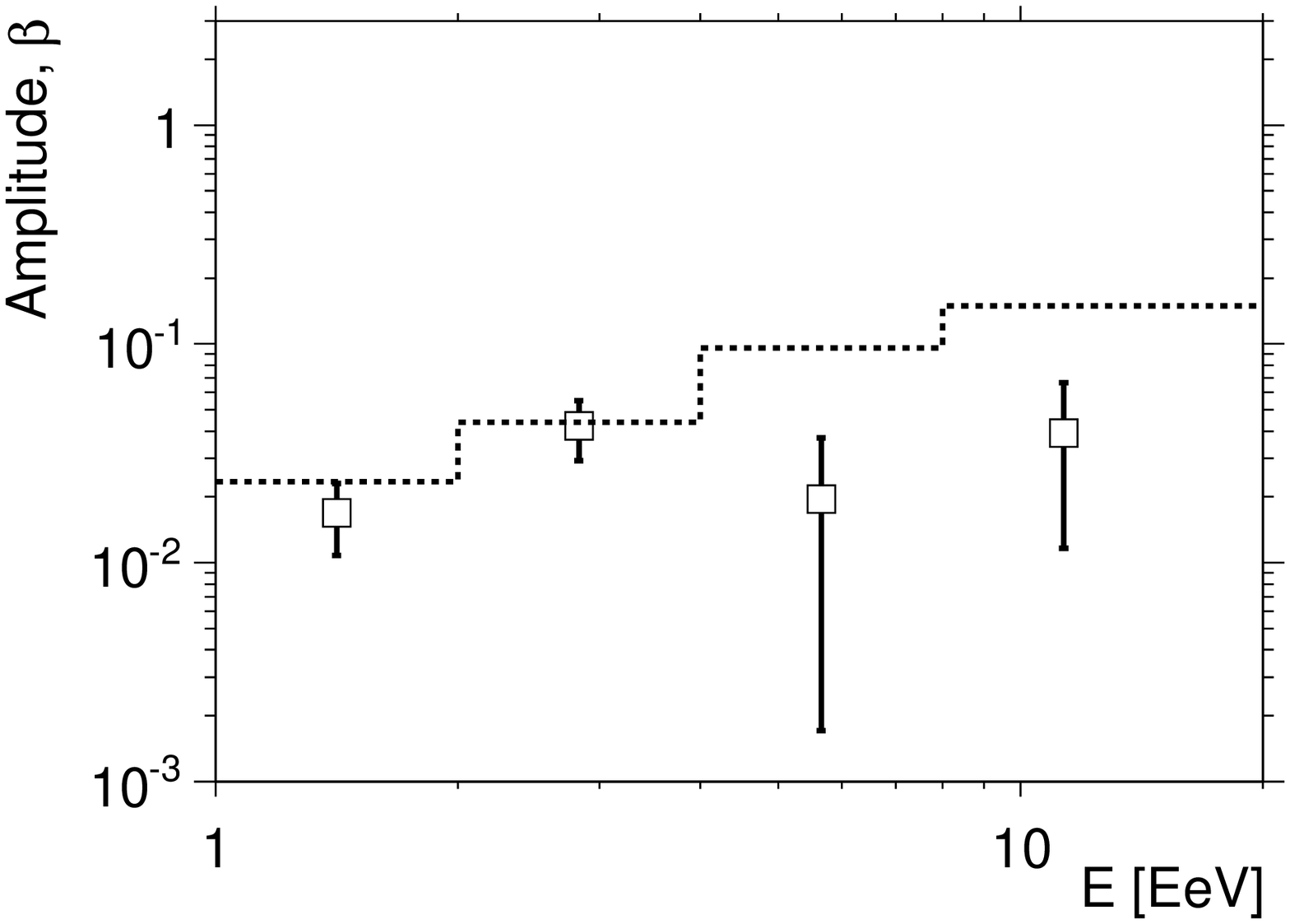}
  \includegraphics[width=7.5cm]{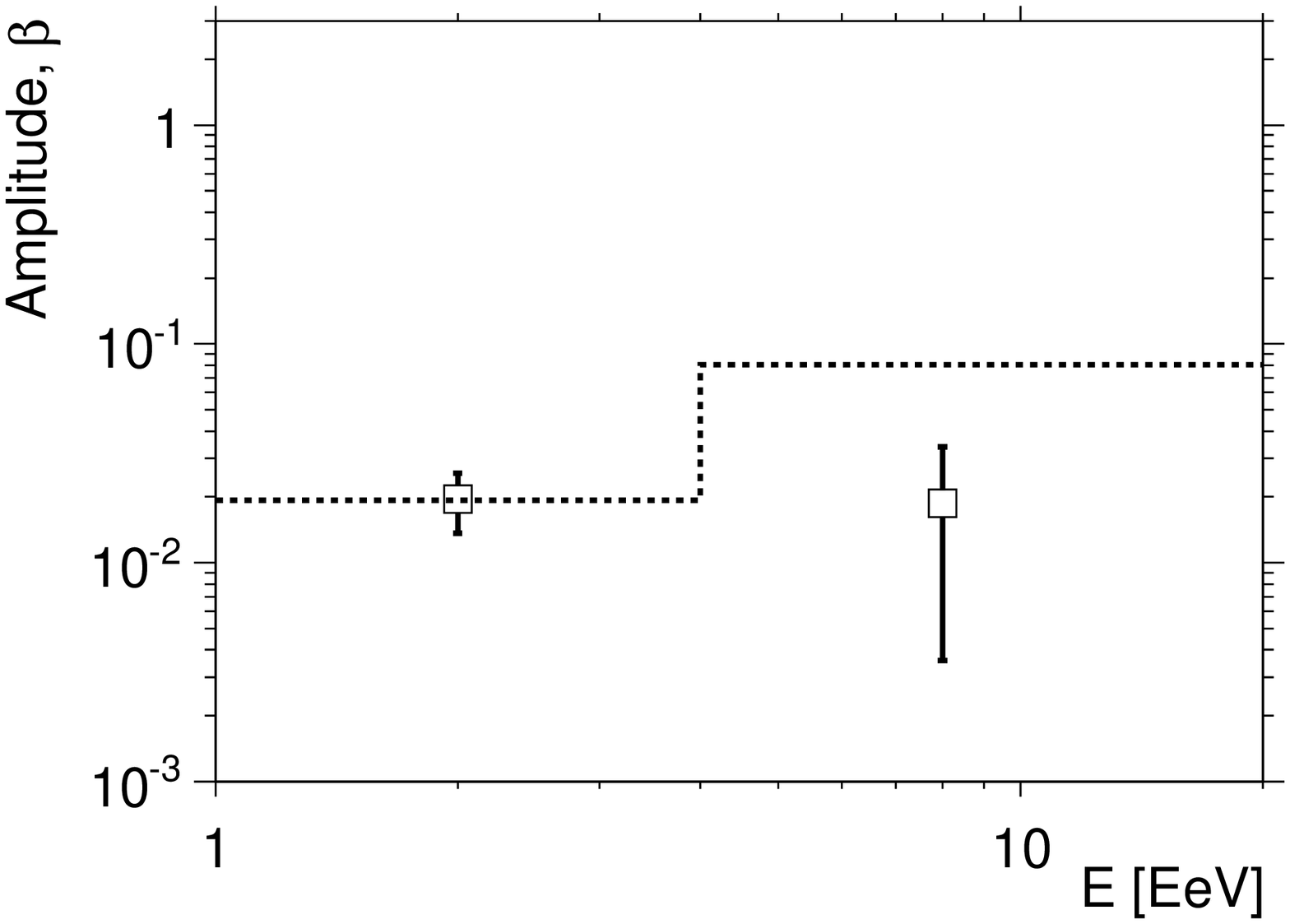}
  \caption{\small{Amplitudes of the dipolar (top) and quadrupolar moments (middle and bottom) 
  as a function of energy using a multipolar reconstruction up to $\ell_{\mathrm{max}}=2$, for two 
  different binnings (left and right). In each panel, the dotted lines stand for the 99\% $C.L.$ upper bounds 
  on the amplitudes that could result from fluctuations of an isotropic distribution.}}
\label{fig:ampquad}
\end{figure}

Assuming now that the angular distribution of cosmic rays is modulated by a dipole \emph{and}
a quadrupole, the intensity $\Phi(\mathbf{n})$ can be parameterised in any direction 
$\mathbf{n}$ as~:
\begin{equation}
\label{eqn:phi-quad}
\Phi(\mathbf{n})=\frac{\Phi_0}{4\pi}~\bigg(1+r~\mathbf{d}\cdot\mathbf{n} +\frac{1}{2}\sum_{i,j}Q_{ij}n_in_j \bigg),
\end{equation}
where $\mathbf{Q}$ is a traceless and symmetric second order tensor. Its five independent
components are determined in a straightforward way from the $\ell=2$ spherical
harmonic coefficients $a_{2m}$. Denoting by $\lambda_+,\lambda_0,\lambda_-$ the 
three eigenvalues of $\mathbf{Q}/2$ ($\lambda_+$ being the highest one and $\lambda_-$ 
the lowest one) and $\mathbf{q_+},\mathbf{q_0},\mathbf{q_-}$ the three 
corresponding unit eigenvectors, the intensity can be parameterised in a more intuitive way as~:
\begin{equation}
\Phi(\mathbf{n})=\frac{\Phi_0}{4\pi}~\bigg(1+r~\mathbf{d}\cdot\mathbf{n} +\lambda_+(\mathbf{q_+}\cdot\mathbf{n})^2 +\lambda_0(\mathbf{q_0}\cdot\mathbf{n})^2 +\lambda_-(\mathbf{q_-}\cdot\mathbf{n})^2 \bigg).
\end{equation}
It is then convenient to define the quadrupole amplitude $\beta$ as~:
\begin{equation}
\beta\equiv\frac{\lambda_+-\lambda_-}{2+\lambda_++\lambda_-}.
\end{equation}
In case of a pure quadrupolar distribution (\textit{i.e.} in the absence of dipole), $\beta$ is 
nothing else but the customary measure of maximal anisotropy contrast~:
\begin{equation}
r=0\Rightarrow \beta=\frac{\lambda_+-\lambda_-}{2+\lambda_++\lambda_-}=\frac{\Phi_{\mathrm{max}}-\Phi_{\mathrm{min}}}{\Phi_{\mathrm{max}}+\Phi_{\mathrm{min}}}.
\end{equation}
Hence, any quadrupolar pattern can be fully described by two amplitudes $(\beta,\lambda_+)$
and three angles~: $(\delta_+,\alpha_+)$ which define the orientation of $\mathbf{q_+}$
and $(\alpha_-)$ which defines the direction of $\mathbf{q_-}$ in the orthogonal plane to 
$\mathbf{q_+}$. The third eigenvector $\mathbf{q_0}$ is orthogonal to $\mathbf{q_+}$ and
$\mathbf{q_-}$, and its corresponding eigenvalue $\lambda_0$ is such that the traceless
condition is satisfied~: $\lambda_++\lambda_-+\lambda_0=0$. Though the probability density
functions of the estimated quadrupole amplitudes $(\overline{\beta},\overline{\lambda}_+)$
can be in principle calculated in the same way as in the case of the estimated dipole
amplitude $(\overline{r})$, expressions are much more complicated to obtain even
semi-analytically and we defer hereafter to Monte-Carlo simulations to tabulate the distributions.

The amplitudes $\overline{r}(E)$, $\overline{\lambda}_+(E)$ and $\overline{\beta}(E)$ are 
shown in Fig.~\ref{fig:ampquad} as functions of energy. Dipole amplitudes are compatible 
with expectations from isotropy. Compared to the results on the dipole obtained in previous section 
for $\ell_{\mathrm{max}}=1$, the sensitivity is now degraded by a factor larger than 2 as 
expected from the dependence of the resolution $\sigma_{\ell m}$ on $\ell_{\mathrm{max}}$ 
(\textit{cf} Eqn.~\ref{eqn:rms-alm}). In the same way as for dipole amplitudes, 
the 99\% $C.L.$ upper bounds on the quadrupole amplitudes that could result from fluctuations 
of an isotropic distribution are indicated by the dashed lines. They correspond to the amplitudes 
$\overline{\lambda}_{+,99}(E)$ and $\overline{\beta}_{99}(E)$ such that the probabilities
$P_{\Lambda_+}(>\overline{\lambda}_{+,99}(E))$ and $P_{B}(>\overline{\beta}_{99}(E))$ 
arising from statistical fluctuations of isotropy are equal to 0.01. Here, both distributions 
$P_{\Lambda_+}$ and $P_{B}$ are sampled from Monte-Carlo simulations. Throughout the
energy scan, there is no evidence for anisotropy. 
The largest deviation from isotropic expectations occurs between 2 and 4~EeV, where both 
amplitudes $\overline{\lambda}_+$ and $\overline{\beta}$ lie just above $\overline{\lambda}_{+99}$ 
and $\overline{\beta}_{99}$.

\section{Additional cross-checks against experimental effects}
\label{systematics}

\subsection{More on the influence of shower size corrections for geomagnetic effects}

Understanding the influence of the shower size corrections for geomagnetic effects is critical 
to get unbiased estimates of anisotropy parameters. Without accounting for these effects, an
increase of the event rate would be observed close to the equatorial South pole with respect
to expectations for isotropy, while a decrease would be observed close to the edge of the 
directional exposure in the equatorial Northern hemisphere. This would result in the 
observation of a fake dipole. A convenient way to exhibit this effect is to separate the 
dipole in two components~: the component of the dipole in the equatorial plane $r_\perp$, 
and the component along the Earth rotation axis, $r_ \parallel$. While $r_\perp$ is 
expected to be affected only by time-dependent effects, $r_ \parallel$ is on the 
other hand the relevant quantity sensitive to time-independent effects such as the geomagnetic 
one.

\begin{table*}[h]
\begin{center}
\begin{tabular}{c|c|c|c|c}
$\Delta E$ [EeV] & $\overline{r}_\perp^{uncorr} [\%]$ & $\overline{r}_\perp [\%]$ & $\overline{r}_\parallel^{uncorr} [\%]$ & $\overline{r}_\parallel [\%]$ \\
\hline
\hline
$1 - 4$ & $0.9\pm0.3$ & $0.9\pm0.3$ & $-2.2\pm0.4$  & $-1.0\pm0.4$  \\
$>4$    & $1.8\pm1.0$ & $2.1\pm1.0$ & $-4.1\pm1.7$  & $-3.0\pm1.7$
\end{tabular}
\caption{\small{Influence of shower size corrections for geomagnetic effects on the component of 
the dipole in the equatorial plane and on the one along the Earth rotation axis.}}
\label{tab:geom}
\end{center}
\end{table*}%

Estimations of  $r_\perp$ and $r_ \parallel$ obtained by accounting or not for 
geomagnetic effects are given in Table~\ref{tab:geom}, in two different energy ranges. 
These estimations are obtained from the recovered $\overline{a}_{1m}$ coefficients~: 
$\overline{r}_\parallel=\sqrt{3}\overline{a}_{10}/\overline{a}_{00}$, and 
$\overline{r}_\perp=[3(\overline{a}_{11}^2+\overline{a}_{1-1}^2]^{0.5}/\overline{a}_{00}$.
It can be seen that the main effect of the geomagnetic corrections is a shift in
$\overline{r}_\parallel$ of about 1.2\%. In the energy range $1\leq E/[\mathrm{EeV}]\leq4$, 
this shift is significant, $\overline{r}_\parallel$ changing from -2.2\% to -1.0\% with an uncertainty 
amounting to 0.4\%. Above 4~EeV, the net correction is of the same order, though the statistical 
uncertainties are larger. In contrast, $\overline{r}_\perp$ remains unchanged in both cases, 
as expected.

\subsection{Eventual energy dependence of the attenuation curve}

In this section, we study to which extent the procedure used to obtain the attenuation
curve in section~\ref{s1000-cic} might influence the determination of the anisotropy parameters.

To convert the shower size into energy, we explained and applied in section~\ref{s1000-cic} 
the constant intensity cut method for showers with $S_{38^\circ}\geq 22~$VEM, that is, just above the 
threshold energy for full efficiency. The value of the parameter $a$ obtained in these conditions 
is consistent within the statistical uncertainties with the one previously reported when applying 
the same constant intensity cut method for showers with $S_{38^\circ}\geq 47~$VEM. Opposite
to this, the value obtained for the coefficient $b$ differs by more than 3 standard
deviations. Such a difference might be expected from both the evolution of the maximum
of the showers and from an eventual change in composition with energy, but it may also
be due to energy and angle-dependent resolutions effects mimicking a real evolution
with energy.

With a different attenuation curve, some events would be reconstructed in the adjacent energy 
intervals in an extent which depends on the change of the attenuation curve with zenith angle. 
For that reason, the determination of anisotropy parameters might be altered by this effect.

Disentangling real evolution of the attenuation curve with energy from resolution effects
is out of the scope of this paper and will be addressed elsewhere. Here, we restrict ourselves
to probe the effect that a real energy dependence would have on the determination of anisotropy 
parameters. To do so, we choose to fit the values of the coefficient $b$ obtained for 
$S_{38^\circ}=22~$VEM and $S_{38^\circ}=47~$VEM through a linear dependence  
with the logarithm of $S_{38^\circ}$. Below and above these values, the behaviour 
of $b(E)$ is obtained by \textit{extrapolating} this energy dependence. In this way, the changes 
in the anisotropy parameters are probed in \textit{extreme} conditions. 

\begin{figure}[!t]
  \centering					 
  \includegraphics[width=9cm]{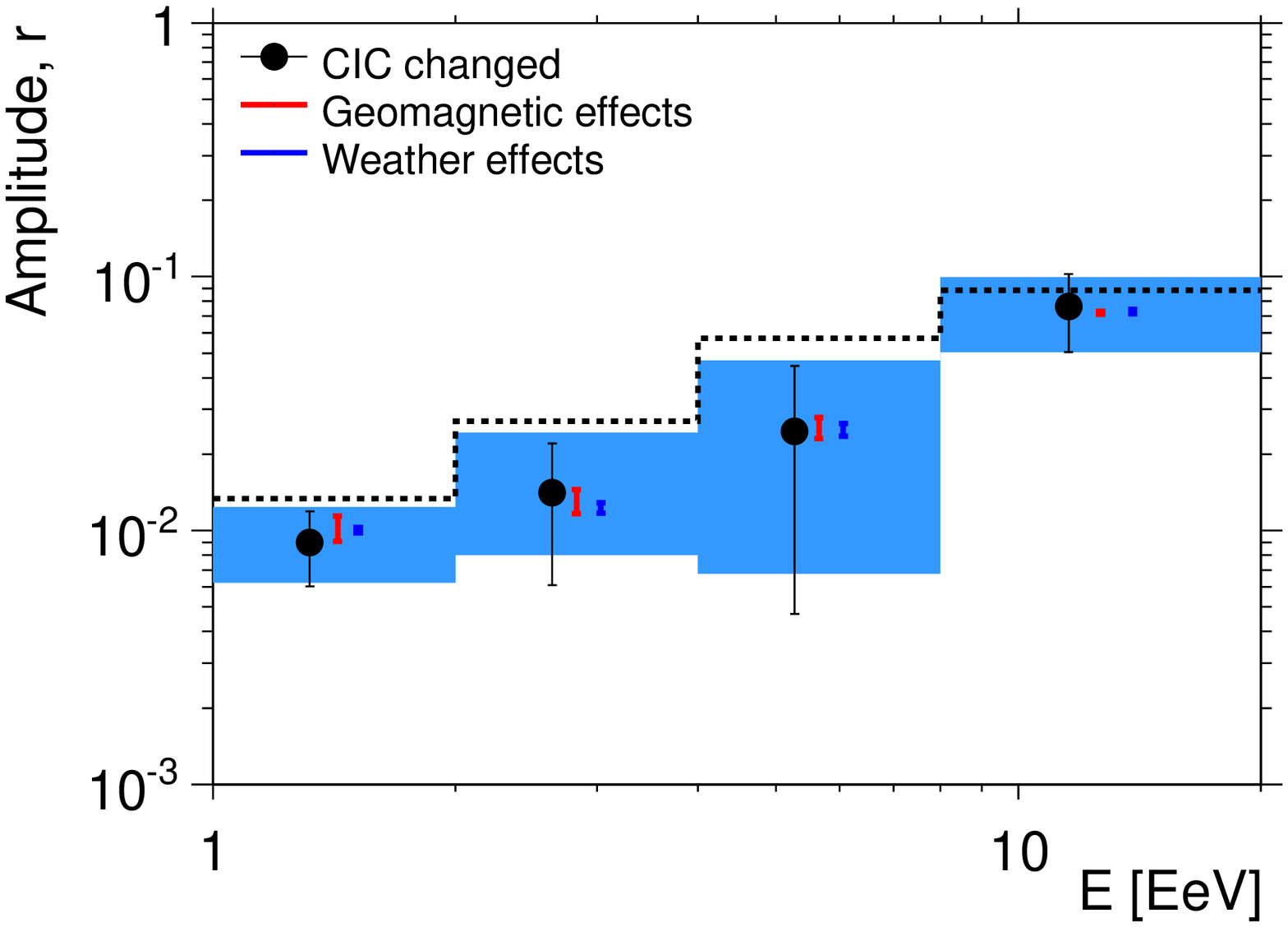}
  \includegraphics[width=7.5cm]{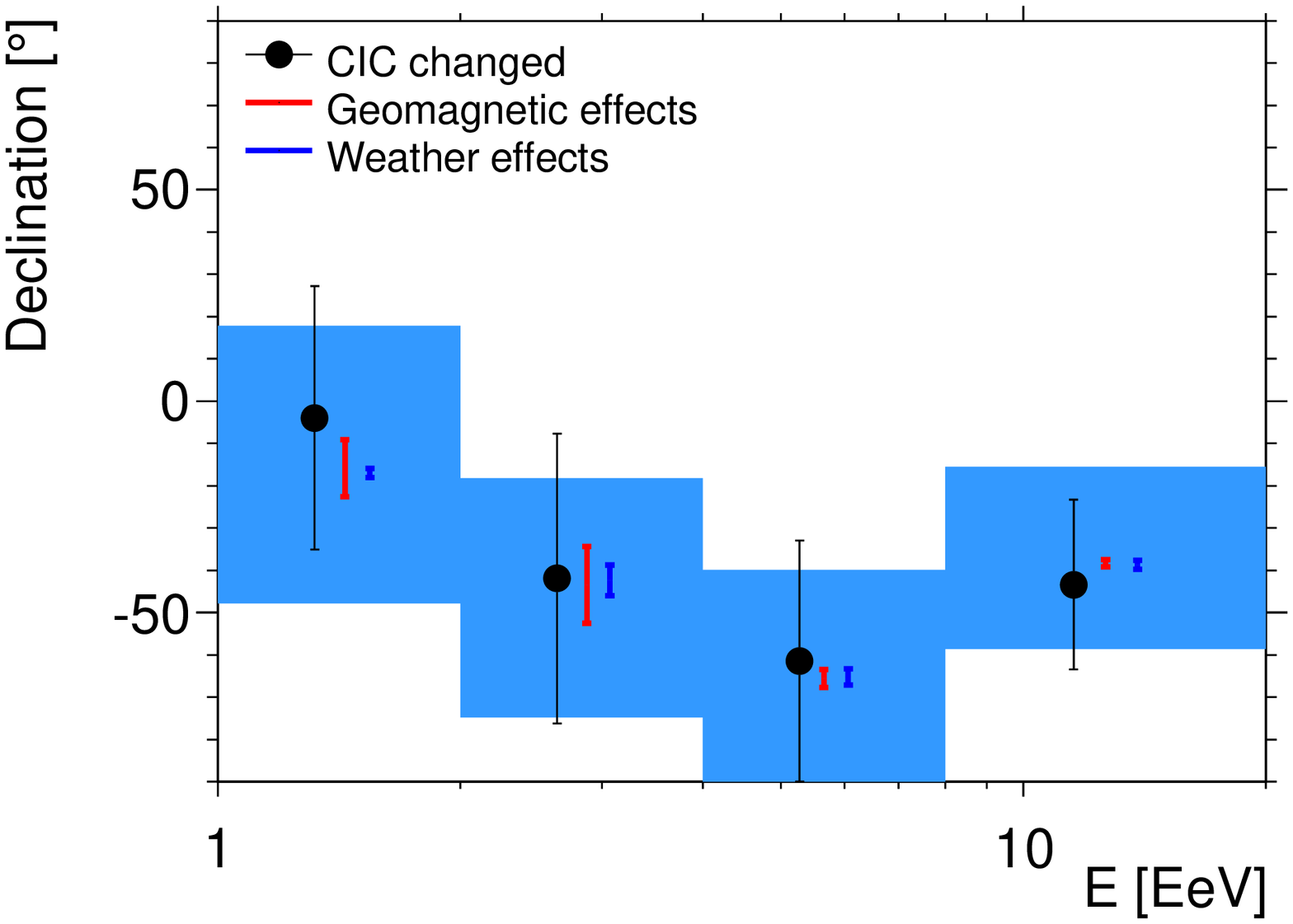}
  \includegraphics[width=7.5cm]{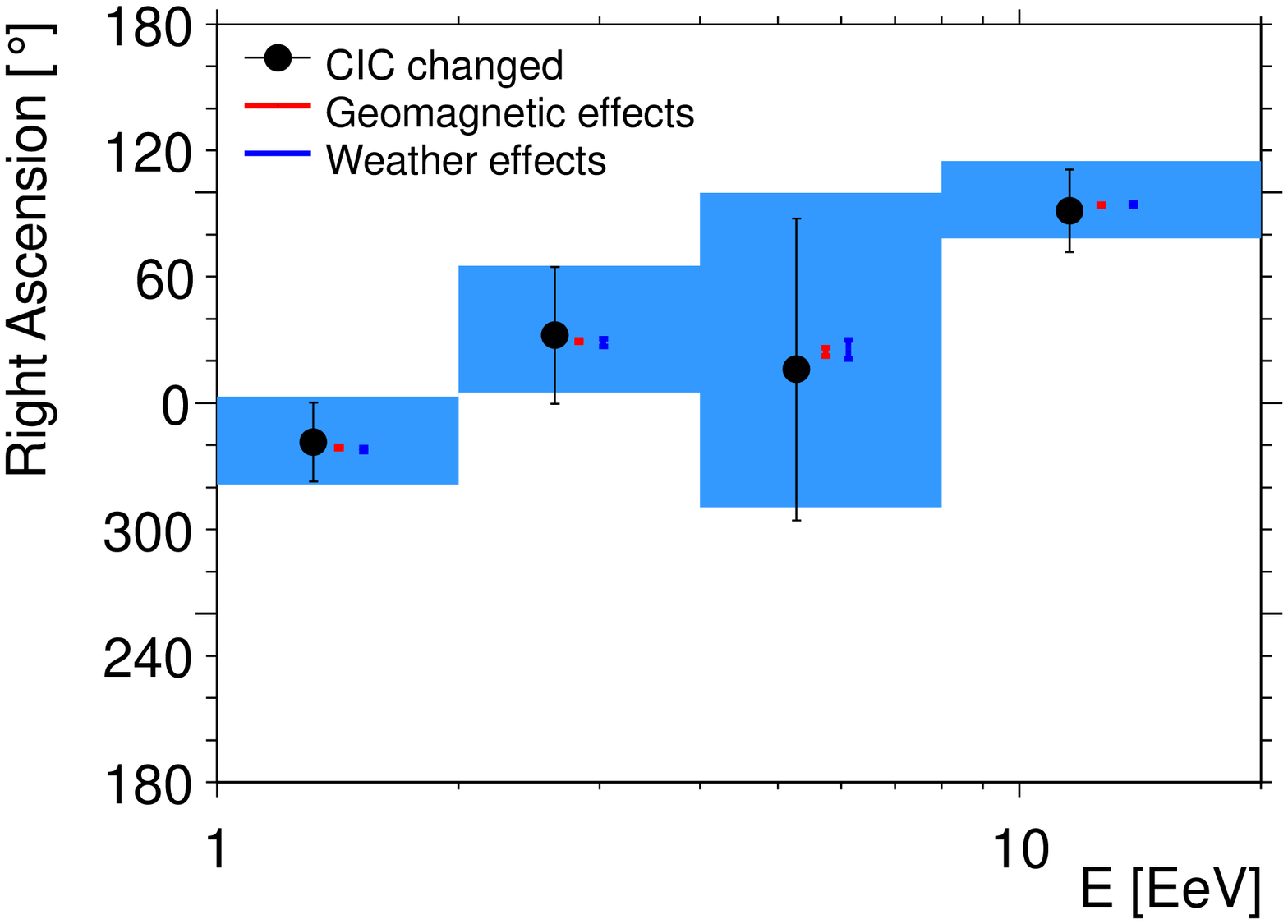}
  \includegraphics[width=7.5cm]{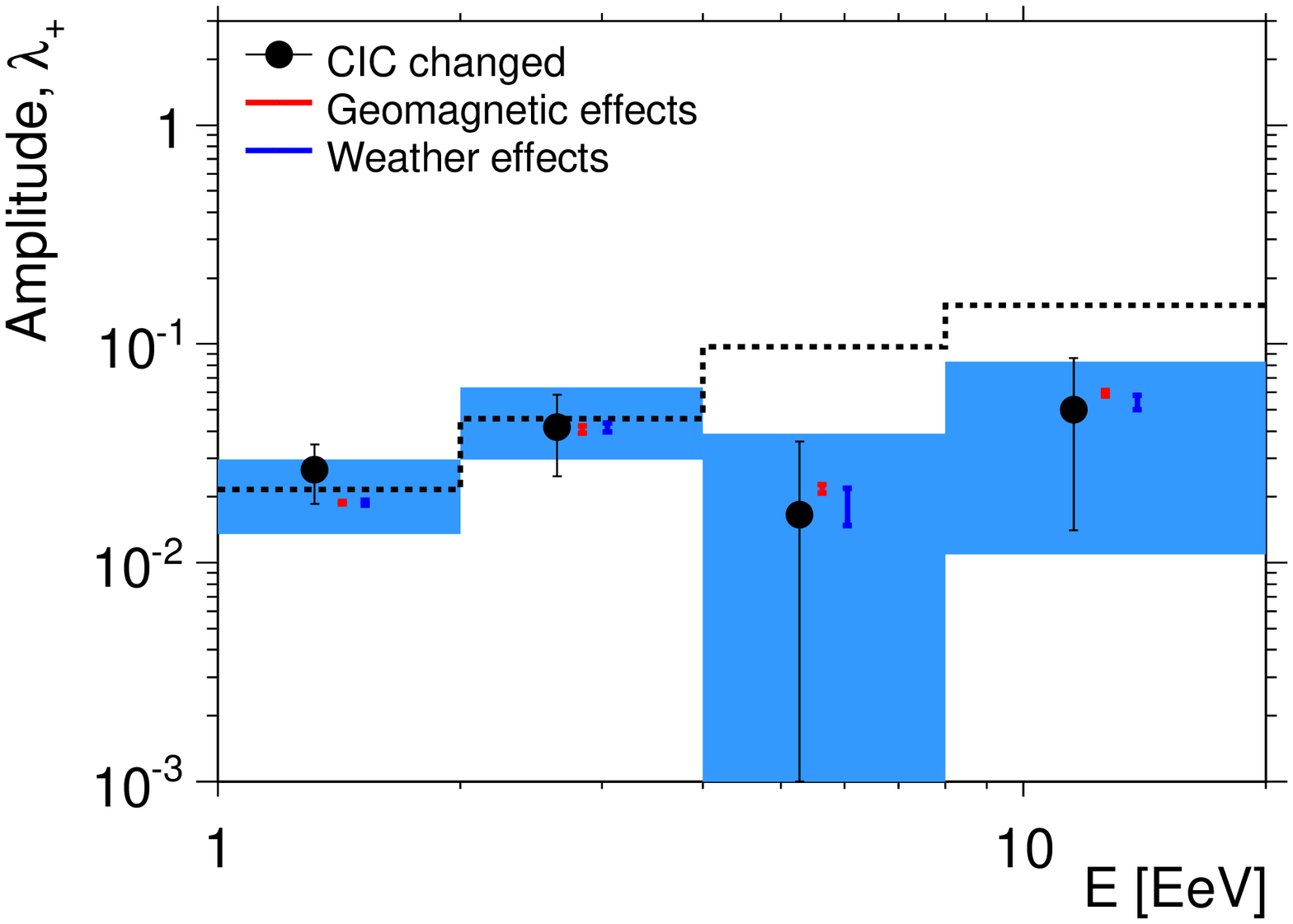}
  \includegraphics[width=7.5cm]{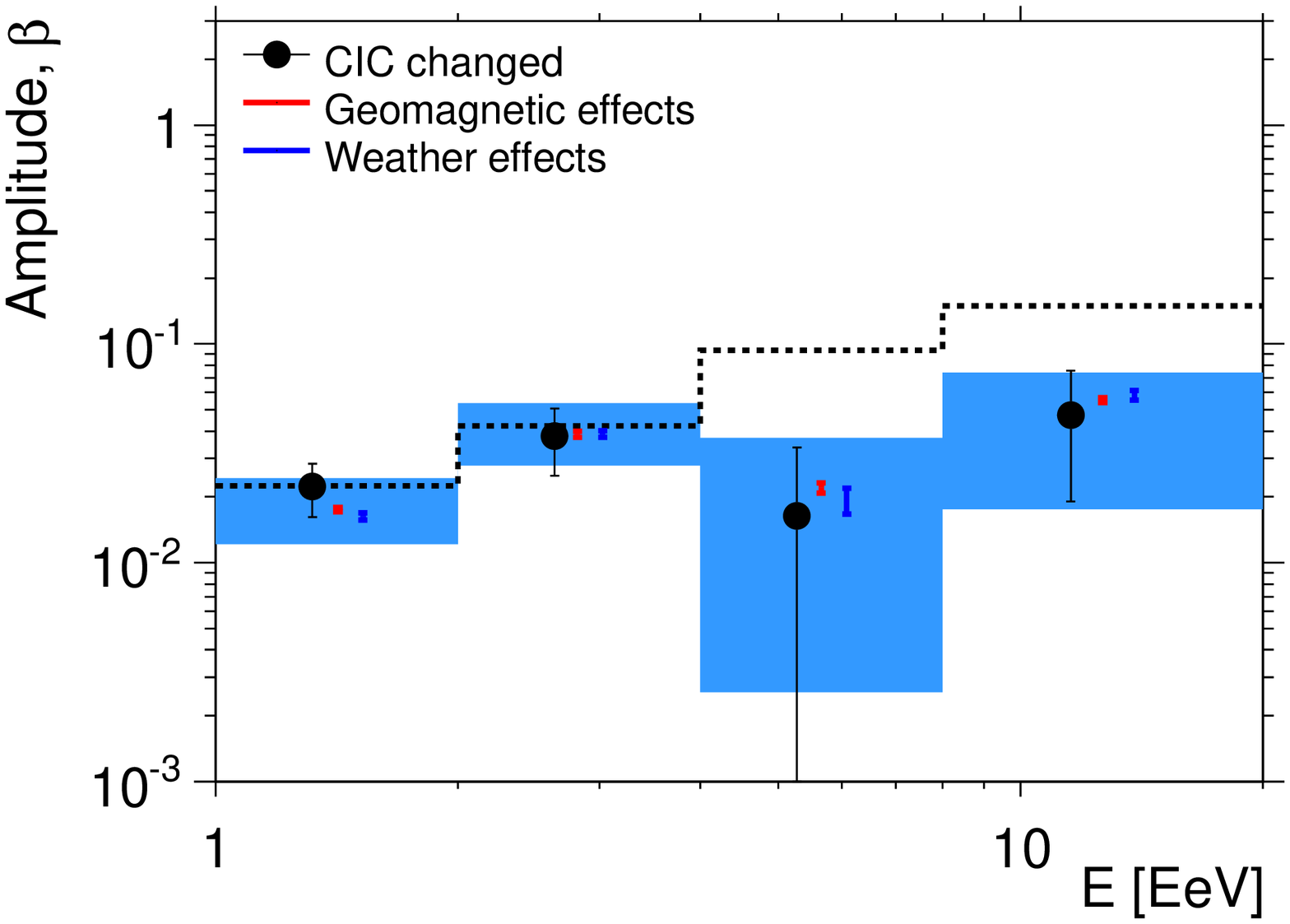}
  \caption{\small{Impact of different sources of systematic uncertainties on the dipole amplitudes (top) and
  the dipole directions and phases (middle) obtained under the assumption $\ell_{\mathrm{max}}=1$, and 
  quadrupole amplitudes (bottom) obtained with $\ell_{\mathrm{max}}=2$, as a function of the energy.
  The blue bands correspond to the results presented in Fig.~\ref{fig:dip} and Fig.~\ref{fig:ampquad}. 
  }}
\label{fig:syst}
\end{figure}

Repeating the whole chain of analysis with this new attenuation curve, it turns out that
the reconstructed dipole parameters are only marginally affected by this change, as illustrated
in the top and middle panels of Fig.~\ref{fig:syst}. Meanwhile, both reconstructed quadrupole amplitudes 
in the energy interval $2\leq E\mathrm{/EeV}\leq 4$ are reduced in such a way that they lie now
just below the 99\% upper bounds for isotropy. Conversely, the amplitudes in the energy interval 
$1\leq E\mathrm{/EeV}\leq 2$ are slightly increased. Below 4~EeV, the determination of the attenuation 
curve thus appears to bring some
systematic uncertainties for determining the quadrupole amplitudes. The two extreme 
extrapolations performed in this analysis (\textit{i.e.} $b$ constant with the energy or linearly
dependent with the logarithm of the energy) allows us to bracket the possible values. 

\subsection{Systematic uncertainties associated to corrections for weather and geomagnetic effects}

In section~\ref{s1000}, we presented the procedure adopted to account for the changes in 
shower size due to weather and geomagnetic effects. Since the coefficients $\alpha_P$, $\alpha_\rho$ 
and $\beta_\rho$ in Eqn.~\ref{sweather} were extracted from real data, they suffer from statistical
uncertainties which may impact in a systematic way the corrections made on $S(1000)$, and
consequently may also impact the anisotropy parameters derived from the data set. Besides,
the determination of $g_1$ and $g_2$ in Eqn.~\ref{sgeom} is based on the simulation of showers.
Both the systematic uncertainties associated to the different interaction models and primary masses
and the statistical uncertainties related to the procedure used to extract $g_1$ and $g_2$
constitute a source of systematic uncertainties on the anisotropy parameters. 

To quantify these systematic uncertainties, we repeated the whole chain of analysis on a large
number of modified data sets. Each modified data set is built by sampling randomly the coefficients
$\alpha_P$, $\alpha_\rho$ and $\beta_\rho$ (or $g_1$ and $g_2$ when dealing
with geomagnetic effects) according to the corresponding uncertainties and correlations between
parameters through the use of a Gaussian probability distribution function. 
For each new set of correction coefficients, new sets of anisotropy parameters are then obtained.
The RMS of each resulting distribution for each anisotropy parameter is the systematic uncertainty
that we assign. Results are shown in Fig.~\ref{fig:syst}, in terms of the dipole and quadrupole
amplitudes as a function of the energy. Balanced against the statistical uncertainties in the 
original analysis (shown by the bands), it is apparent that both sources of systematic uncertainties 
have a negligible impact on each reconstructed anisotropy amplitude.

\section{Upper limits and discussion}
\label{discussion}

\begin{figure}[!t]
  \centering					 
  \includegraphics[width=7.5cm]{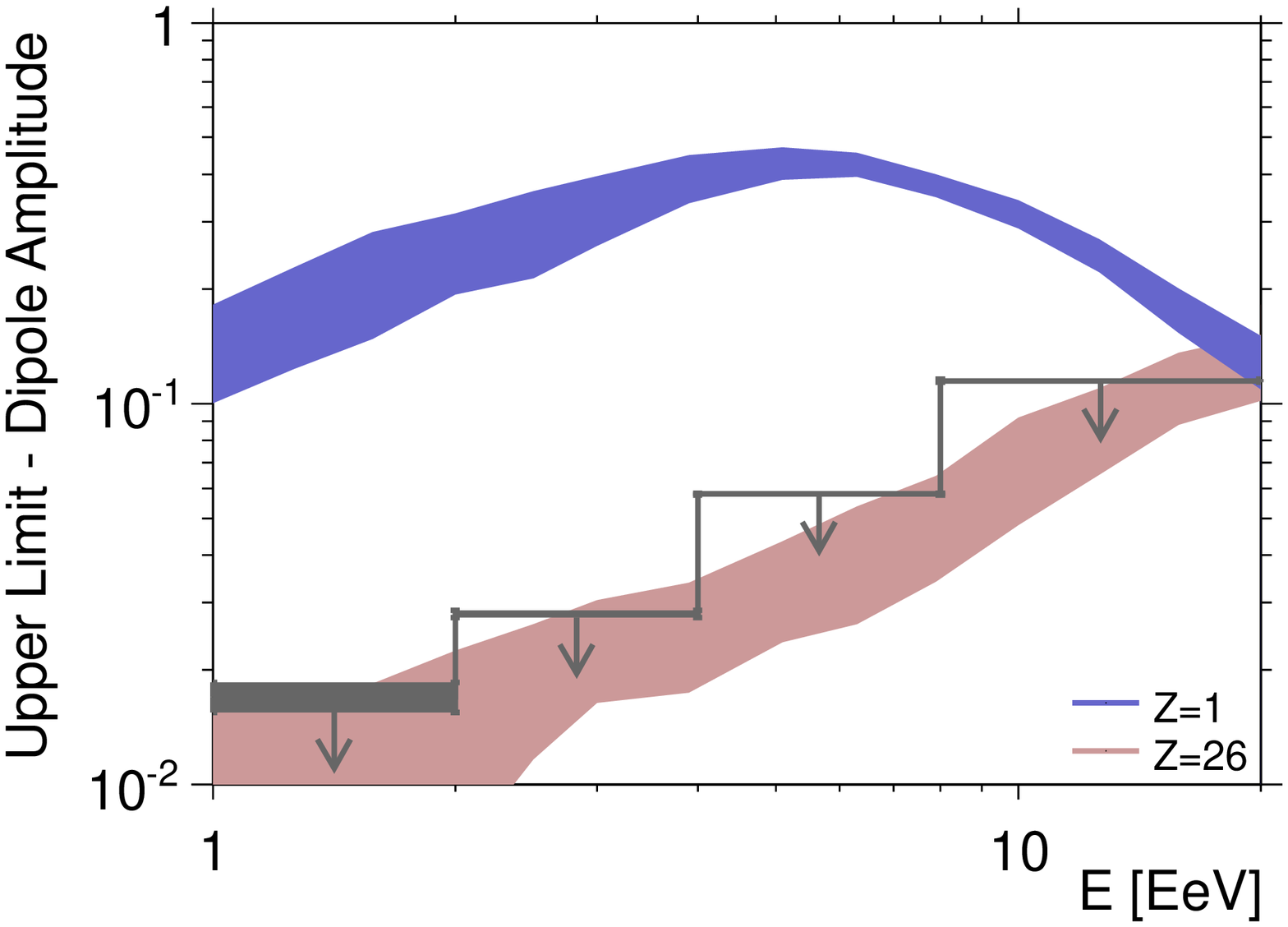}
  \includegraphics[width=7.5cm]{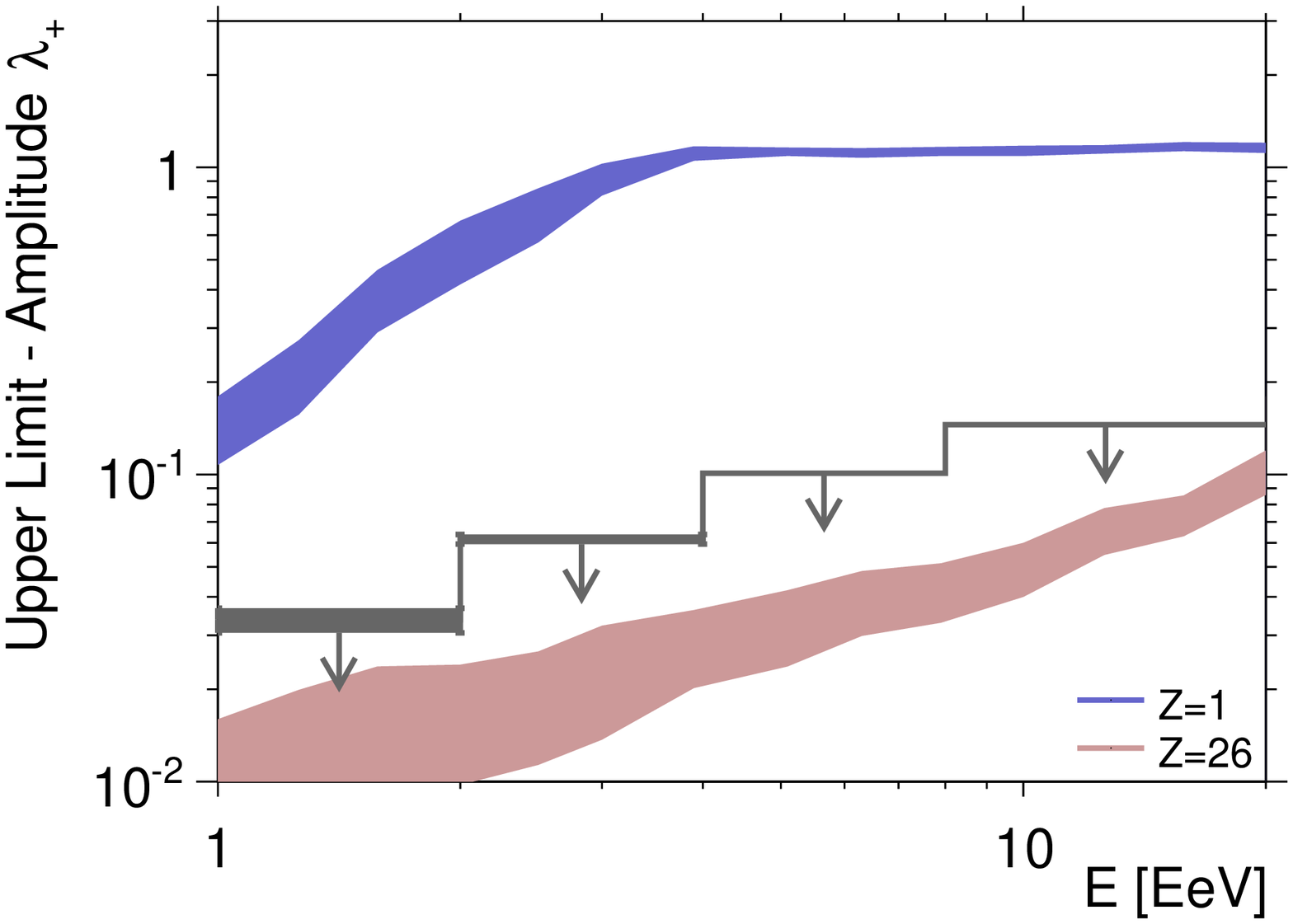}
  \caption{\small{99\% $C.L.$ upper limits on dipole and quadrupole amplitudes as a function of the energy.
  Some generic anisotropy expectations from stationary galactic sources distributed in the disk are
  also shown, for various assumptions on the cosmic ray composition. The fluctuations of the amplitudes
  due to the stochastic nature of the turbulent component of the magnetic field are sampled from different 
  simulation data sets and are shown by the bands (see text).}}
\label{fig:UL}
\end{figure}

From the analyses reported in section~\ref{analysis}, upper limits on dipole and quadrupole
amplitudes can be derived at 99\% $C.L.$ (see appendices C and D). All relevant results are  
summarised in Table~\ref{tab:dip} and Table~\ref{tab:quad}. The upper limits are also shown 
in Fig.~\ref{fig:UL} accounting for the systematic uncertainties discussed in the previous section~:
in the two last energy bins, the upper limits are quite insensitive to the systematic uncertainties 
because all amplitudes lie well within the background noise. 

We illustrate below the astrophysical interest of these upper limits by calculating the anisotropy 
amplitudes expected in a toy scenario in which sources of
EeV-cosmic rays are stationary, densely and uniformly distributed in the galactic disk, and emit 
particles in all directions.

\begin{table*}[h]
\begin{center}
\begin{tabular}{c|c|c|c|c|c}
$\Delta E$ [EeV] & $N$ & $\overline{r}$ [\%] & $\overline{\delta} [^\circ]$  & $\overline{\alpha} [^\circ]$  & UL [\%] \\
\hline
\hline
$1 - 2$ & 360132 & $1.0 \pm 0.4$ & $-15 \pm 32$  & $342\pm 20$ & 1.5 \\
$2 - 4$ & 88042 & $1.6 \pm 0.8$ & $-46\pm 28$  & $35\pm 30$ & 2.8  \\
$4 - 8$ & 19794 & $2.7\pm 2.0$ & $-69\pm 30$  & $25\pm 74$ &  5.8 \\
$>8$ & 8364 & $7.5\pm 2.5$ & $-37\pm 21$  & $96\pm 18$ &  11.4 
\end{tabular}
\caption{\small{Summary of the dipolar analysis ($\ell_{\mathrm{max}}=1$) reported in section~\ref{dipole}, 
together with the derived 99\% $C.L.$ upper limits (UL) on the amplitudes.}}
\label{tab:dip}
\end{center}
\end{table*}%

\begin{table*}[h]
\begin{center}
\begin{tabular}{c|c|c|c|c}
$\Delta E$ [EeV] & $\overline{\lambda}_+$ [\%] & $\overline{\beta}$ [\%] & UL ($\lambda_+$) [\%]  & UL ($\beta$) [\%] \\
\hline
\hline
$1 - 2$ & $2.0 \pm 0.7$  & $1.7 \pm 0.6$ & $3.0$  & $2.9$ \\
$2 - 4$ & $5.0 \pm 1.7$ & $4.2 \pm 1.3$ & $6.3$  & $6.1$  \\
$4 - 8$ & $1.6 \pm 2.0$ & $1.9\pm 1.8$ & $10.0$  & $9.4$  \\
$>8$ & $4.0 \pm 3.4$  & $3.9\pm 2.7$ & $14.5$  & $13.8$ 
\end{tabular}
\caption{\small{Summary of the quadrupolar analysis ($\ell_{\mathrm{max}}=2$) reported in section~\ref{quadrupole}, 
together with the derived 99\% $C.L.$ upper limits (UL) on the amplitudes.}}
\label{tab:quad}
\end{center}
\end{table*}%

Both the strength and the structure of the magnetic field in the Galaxy, known only approximately, 
play a crucial role in the propagation of cosmic rays. The field is thought to contain a large 
scale regular component and a small scale turbulent one, both having a local strength of a few microgauss 
(see \textit{e.g.}~\citep{Beck2001}). While the turbulent component dominates in strength by a factor 
of a few, the regular component imprints dominant drift motions as soon as the Larmor radius of
cosmic rays is larger than the maximal scale of the turbulences (thought to be in the range 10-100~pc). 
We adopt in the following a recent parameterisation of the regular component obtained by fitting 
model field geometries to Faraday rotation measures of extragalactic radio sources and polarised
synchrotron emission~\citep{Pshirkov2011}.
It consists in two different components~: a disk field and a halo field. The disk field is symmetric with
respect to the galactic plane, and is described by the widely-used logarithmic spiral model with
reversal direction of the field in two different arms (the so-called \textit{BSS-model}). The halo field
is anti-symmetric with respect to the galactic plane and purely toroidal. The detailed parameterisation
is given in Ref.~\citep{Pshirkov2011} (with the set of parameters reported in Table~3). In addition to
the regular component, a turbulent field is generated according to a Kolmogorov power spectrum
and is pre-computed on a three dimensional grid periodically repeated in space. The size of the 
grid is taken as 100~pc, so as the maximal scale of turbulences, and the strength of the turbulent 
component is taken as three times the strength of the regular one.

To describe the propagation of cosmic rays with energies $E\geq 1$~EeV in such 
a magnetic field, the direct integration of trajectories is the most appropriate tool. Performing the
forward tracking of particles from galactic sources and recording those particles which 
cross the Earth is however not feasible within a reasonable computing time. So, to obtain 
the anisotropy of cosmic rays emitted from sources uniformly distributed in a disk with a radius
of 20~kpc from the galactic centre and with a height of $\pm$ 100~pc, we adopt a method first proposed 
in Ref.~\citep{Thielheim1968} and then widely used in the literature.
It consists in back tracking anti-particles with random directions from the Earth to outside the Galaxy. 
Each test particle \textit{probes} the total luminosity along the path of propagation from each
direction as seen from the Earth. For \textit{stationary sources emitting cosmic rays in all directions},
the flux expected in a given sampled direction is then proportional to the time spent in the source region 
by the test particles arriving from that direction.

The amplitudes of anisotropy obviously depend on the rigidity $E/Z$ of the cosmic rays, with $Z$ the
electric charge of the particles. Since we only aim at illustrating the upper limits, we consider two 
extreme single primaries~: protons and iron nuclei. In the energy range 
$1\leq E/\mathrm{EeV}\leq 20$, it is unlikely that our measurements on the average position in the 
atmosphere of the shower maximum and the corresponding RMS can be reproduced with a single 
primary~\citep{AugerPRL2010}. As well, in the scenario explored here and for a single primary, the 
energy spectrum is expected to reveal a \textit{hardening} in this energy range, whose origin is different
from the one expected if the ankle marks the cross-over between galactic and extragalactic cosmic
rays~\citep{Linsley1963} or if it marks the distortion of a proton-dominated extragalactic spectrum 
due to $e^+/e^-$ pair production of protons with the photons of the cosmic microwave 
background~\citep{Hillas1967,Blumenthal1970,Berezinsky2006,Berezinsky2004}.
For a given configuration of the magnetic field, the exact energy at which this hardening occurs depends 
on the electric charge of the cosmic rays. This is because the average time spent in the source region 
first decreases as $\simeq E^{-1}$ and then tends to the constant free escape time as a consequence 
of the direct escape from the Galaxy. The hardening with $\Delta\gamma\simeq0.6$ observed at 4~EeV 
in our measurements of the energy spectrum is not compatible with the one expected in this scenario 
($\Delta\gamma\simeq1$). Nevertheless, the calculation of dipole and quadrupole amplitudes for single 
primaries is useful to probe the allowed contribution of each primary as a function of the energy.

The dipole $r$ and quadrupole $\lambda_+$ amplitudes obtained for several energy values covering 
the range $1\leq E/\mathrm{EeV}\leq 20$ are shown in Fig.~\ref{fig:UL}. To probe unambiguously 
amplitudes down to the percent level, it is necessary to generate simulated event sets with 
$\simeq 5~10^5$ test particles. Such a number of simulated events allows us to shrink statistical 
uncertainties on amplitudes at the $0.5$\% level. Meanwhile, there is an intrinsic variance in the model 
for each anisotropy parameter due to the stochastic nature of the turbulent component of the magnetic field. 
This variance is estimated through the simulation of 20 sets of $5~10^5$ test particles, where
the configuration of the turbulent component is frozen in each set. The RMS of the amplitudes sampled 
in this way is shown by the bands in Fig.~\ref{fig:UL}. While the dipole amplitude steadily increases 
for iron nuclei, this is not the case any longer for protons around the ankle energy. This is 
because we explore a source region uniformly distributed in the disk. Consequently, the image of 
the galactic plane appears less distorted by the magnetic field with increasing energy.
This gives rise to an important quadrupolar moment which actually turns out to be the main feature 
of the anisotropy at large scale~\footnote{This feature would remain in the case of a radial distribution
of sources following the matter in the Galaxy, though the dipole amplitude would steadily 
increase above the ankle energy.}. 

The dipole and quadrupole $\lambda_+$ amplitudes obtained here depend on the model used 
to describe the galactic magnetic field. We note that recently, a new model was given in 
Ref.~\citep{Farrar2012}, providing improved fits to Faraday rotation measures of extragalactic radio 
sources and polarised synchrotron emission observations. However, we tested at a few energies that 
the results obtained are qualitatively in agreement with the ones presented in Fig.~\ref{fig:UL}. 
Similar conclusions were given in Ref.~\citep{Giacinti2011}, where more systematic studies can be 
found in terms of the field strength and geometry.

Around $1~$EeV, there are indications that the cosmic ray composition includes a significant light
component from various measurements of the depth of shower maximum 
$X_{max}$~\citep{AugerPRL2010,HiResPRL2010,TAAPS2011}. It is apparent that amplitudes 
derived for protons largely stand above the allowed limits. Consequently, unless the strength of the 
magnetic field is much higher than in the picture used here, the upper limits derived in this analysis 
exclude that the light component of cosmic rays comes from galactic stationary sources densely 
distributed in the galactic disk and emitting in all directions. This is in agreement with the absence of 
any detectable point-like sources above 1~EeV that would be indicative of a flux of neutrons produced
by EeV-protons through mainly pion-producing interactions in the source 
environments~\citep{Auger2012submit}. On the other hand, if the 
cosmic ray composition around $1~$EeV results from a mixture containing a large fraction of iron 
nuclei of galactic origin, upper limits can still be respected, or alternatively a light component of 
extragalactic origin would be allowed. Future measurements of composition 
below $1~$EeV will come from the low energy extension HEAT now available at the Pierre Auger 
Observatory~\citep{AugerHEAT}. 
Combining these measurements with large scale anisotropy ones will then allow us to further 
understand the origin of cosmic rays at energies less than 4~EeV.

\section{Summary}
\label{summary}

For the first time, a thorough search for large scale anisotropies as a function of both the declination 
and the right ascension in the distribution of arrival directions of cosmic rays detected above 
$1$~EeV at the Pierre Auger Observatory has been presented. With respect to the 
traditional search in right ascension only, this search requires the control of additional systematic
effects affecting both the exposure of the sky and the counting rate of events in local angles.
All these effects were carefully accounted for and presented in sections~\ref{s1000} 
and~\ref{exposure}. No significant deviation from isotropy is revealed within the systematic 
uncertainties, although the consistency in the dipole phases may be indicative of a genuine signal 
whose amplitude is at the level of the statistical noise. The sensitivity accumulated so far to 
dipole and quadrupole amplitudes allows us to challenge an origin of cosmic rays from stationary 
galactic sources densely distributed in the galactic disk and emitting predominantly light particles 
in all directions. 

Future work will profit from both the increased statistics and the lower energy threshold that 
is now available at the Pierre Auger Observatory~\citep{AugerHEAT,AugerAmiga}. This will 
provide further constraints helping to understand the origin of cosmic rays in the energy range
$0.1<E/\mathrm{EeV}<10$.

\section*{Acknowledgements}

The successful installation, commissioning, and operation of the Pierre Auger Observatory
would not have been possible without the strong commitment and effort
from the technical and administrative staff in Malarg\"ue.

We are very grateful to the following agencies and organizations for financial support: 
Comisi\'on Nacional de Energ\'ia At\'omica, 
Fundaci\'on Antorchas,
Gobierno De La Provincia de Mendoza, 
Municipalidad de Malarg\"ue,
NDM Holdings and Valle Las Le\~nas, in gratitude for their continuing
cooperation over land access, Argentina; 
the Australian Research Council;
Conselho Nacional de Desenvolvimento Cient\'ifico e Tecnol\'ogico \linebreak (CNPq),
Financiadora de Estudos e Projetos (FINEP),
Funda\c{c}\~ao de Amparo \`a Pesquisa do Estado de Rio de Janeiro (FAPERJ),
Funda\c{c}\~ao de Amparo \`a Pesquisa do Estado de S\~ao Paulo (FAPESP),
Minist\'erio de Ci\^{e}ncia e Tecnologia (MCT), Brazil;
AVCR AV0Z10100502 and AV0Z10100522, GAAV KJB100100904, MSMT-CR LA08016,
LG11044, MEB111003, MSM0021620859, LA08015 and TACR TA01010517, Czech Republic;
Centre de Calcul IN2P3/CNRS, 
Centre National de la Recherche Scientifique (CNRS),
Conseil R\'egional Ile-de-France,
D\'epartement  Physique Nucl\'eaire et Corpusculaire (PNC-IN2P3/CNRS),
D\'epartement Sciences de l'Univers (SDU-INSU/CNRS), France;
Bundesministerium f\"ur Bildung und Forschung (BMBF),
Deutsche Forschungsgemeinschaft (DFG),
Finanzministerium Baden-W\"urttemberg,
Helmholtz-Gemeinschaft Deutscher Forschungszentren (HGF),
Ministerium f\"ur Wissenschaft und Forschung, Nordrhein-Westfalen,
Ministerium f\"ur Wissenschaft, Forschung und Kunst, Baden-W\"urttemberg, Germany; 
Istituto \linebreak  Nazionale di Fisica Nucleare (INFN),
Ministero dell'Istruzione, dell'Universit\`a e della Ricerca (MIUR), Italy;
Consejo Nacional de Ciencia y Tecnolog\'ia (CONACYT), Mexico;
Ministerie van Onderwijs, Cultuur en Wetenschap,
Nederlandse Organisatie voor Wetenschappelijk Onderzoek (NWO),
Stichting voor Fundamenteel Onderzoek der Materie (FOM), Netherlands;
Ministry of Science and Higher Education,
Grant Nos. N N202 200239 and N N202 207238, Poland;
Portuguese national funds and FEDER funds within COMPETE - Programa Operacional Factores de Competitividade through 
Funda\c{c}\~ao para a Ci\^{e}ncia e a Tecnologia, Portugal;
Romanian Authority for Scientific Reseach, UEFICDI,
Ctr.Nr.1/ASPERA2 ERA-NET, Romania; 
Ministry for Higher Education, Science, and Technology,
Slovenian Research Agency, Slovenia;
Comunidad de Madrid, 
FEDER funds, 
Ministerio de Ciencia e Innovaci\'on and Consolider-Ingenio 2010 (CPAN),
Xunta de Galicia, Spain;
Science and Technology Facilities Council, United Kingdom;
Department of Energy, Contract Nos. DE-AC02-07CH11359, DE-FR02-04ER41300,
National Science Foundation, Grant No. 0450696,
The Grainger Foundation USA; 
NAFOSTED, Vietnam;
Marie Curie-IRSES/EPLANET, European Particle Physics Latin American Network, 
European Union 7th Framework Program, Grant No. PIRSES-2009-GA-246806; 
and UNESCO.

\section*{Appendix A~: Large scale anisotropies in local coordinates}
\label{appendixA}

\begin{figure}[!h]
  \centering					 
  \includegraphics[width=7.5cm]{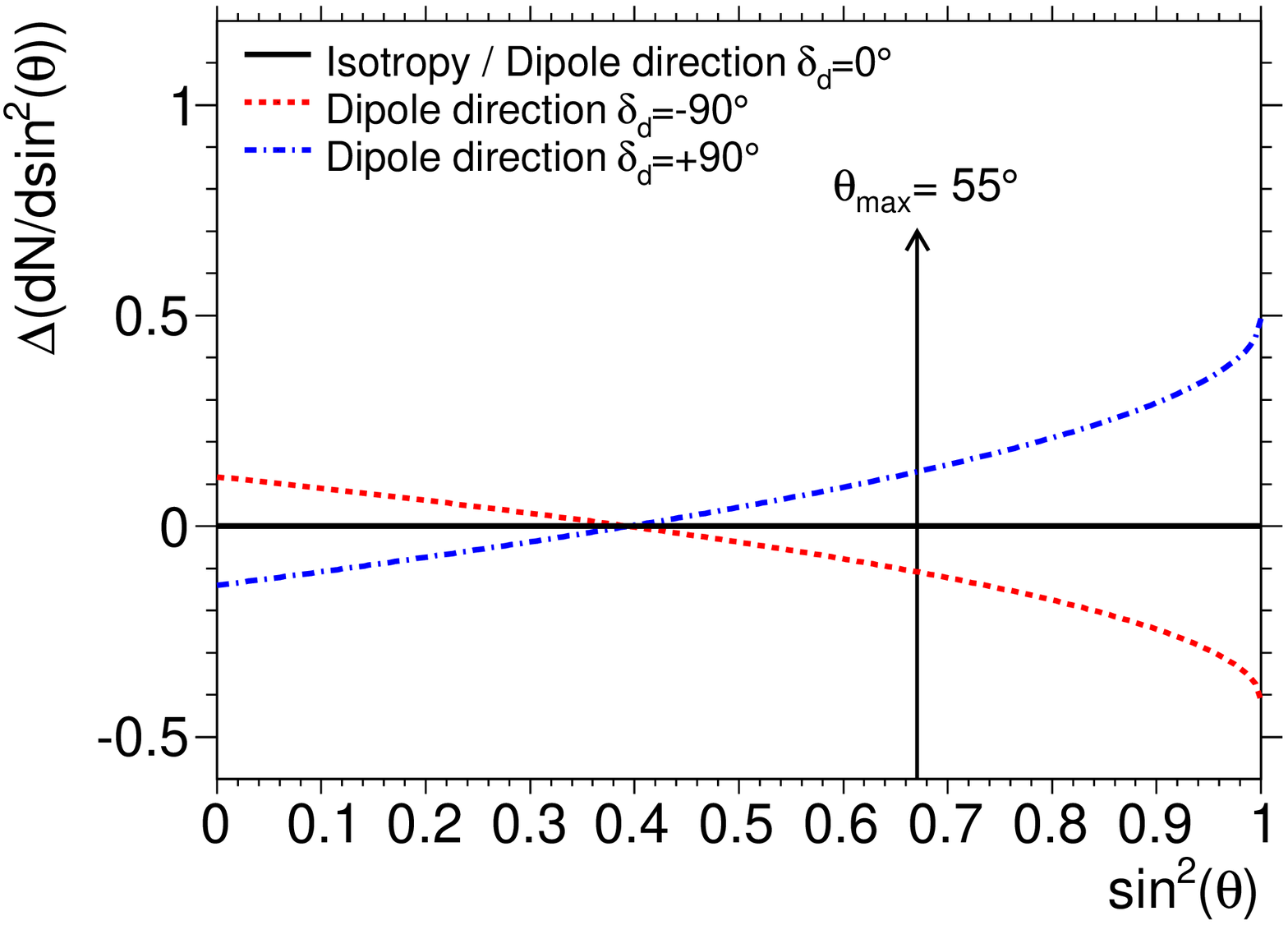}
  \includegraphics[width=7.5cm]{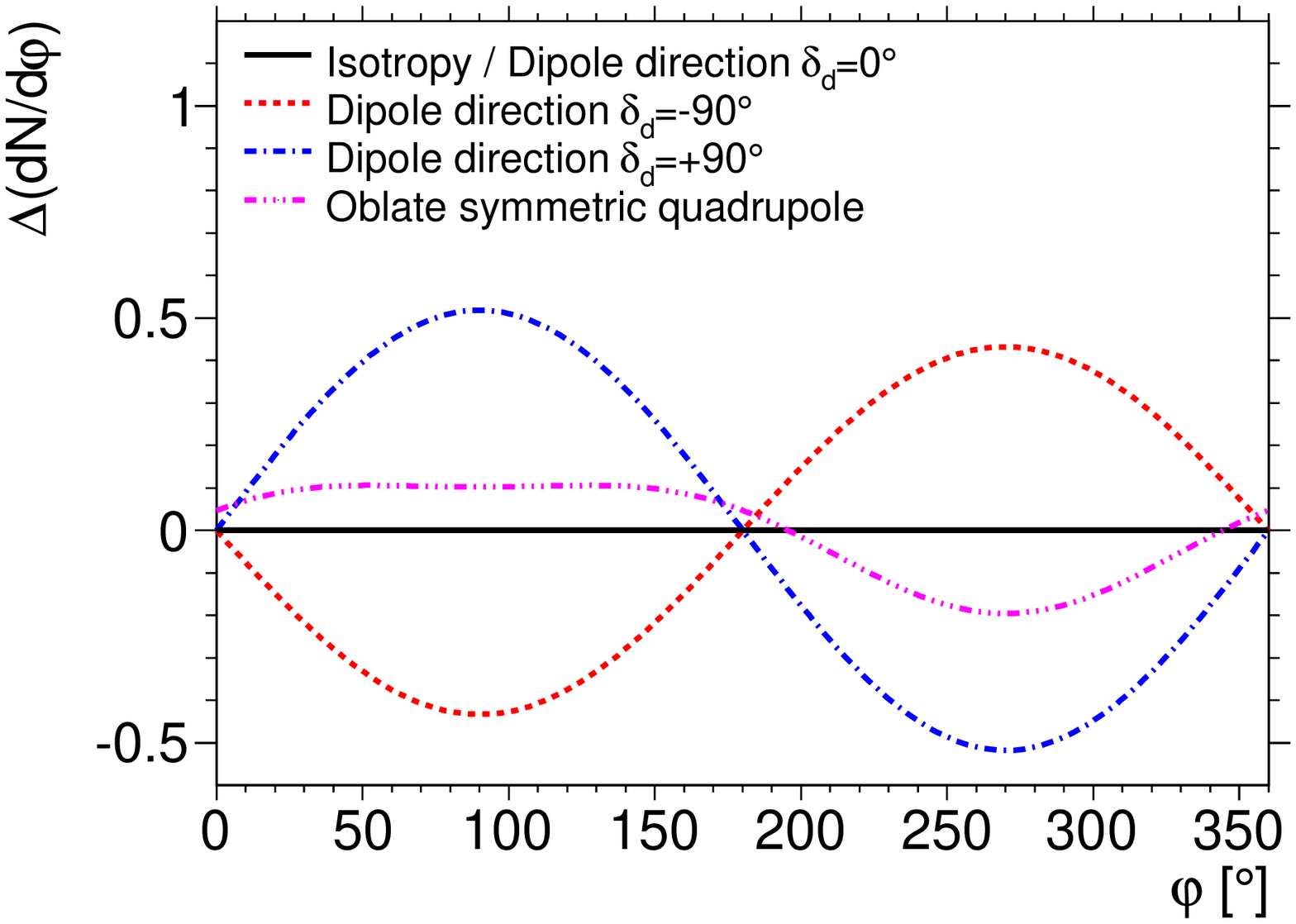}
  \caption{\small{Effect of large scale anisotropies in local coordinates (left~: as a function of $\sin^2{\theta}$,
  right~: as a function of $\varphi$) for an observer located at the Earth latitude 
  $\ell_{\mathrm{site}}=-35.2^\circ$ of the Pierre Auger Observatory.}}
\label{fig:localangles}
\end{figure}

To study the angular distribution in local coordinates for different anisotropic angular distributions
$\Phi(\alpha,\delta)$ in celestial coordinates, we restrict ourselves, without loss of generalities, to 
the case of full detection efficiency ($\epsilon(\theta,\varphi,E)=1$). 
Then, the instantaneous arrival direction distribution in local coordinates reads~:
\begin{equation}
\label{eqn:d3N}
\frac{\mathrm{d}^3N}{\mathrm{d}\theta \mathrm{d}\varphi \mathrm{d}\alpha^0} \propto \sin{\theta}~\cos{\theta}~\Phi(\theta,\varphi,\alpha^0).
\end{equation}
$\Phi(\theta,\varphi,\alpha^0)$ is the underlying angular distribution of cosmic rays, expressed
in local coordinates. In case of isotropy, $\Phi$ is constant so that once integrated over $\varphi$ 
and $\alpha^0$, the arrival direction distribution is such that $\mathrm{d}N/\mathrm{d}\sin^2{\theta}$ 
is also constant. On the other hand, in case of a dipolar distribution for instance, $\Phi$ is proportional to 
$1+r\mathbf{d}(\theta,\varphi,\alpha^0)\cdot \mathbf{n}(\theta,\varphi)$, where $\mathbf{n}$ is here
a unit vector in local coordinates, and $\mathbf{d}$ the dipole unit vector pointing towards 
$(\alpha_d,\delta_d)$ and expressed in local coordinates by means of Eqn.~\ref{eqn:theta-phi}. 
To quantify the distortions induced by a dipole in the $\mathrm{d}N/\mathrm{d}\sin^2{\theta}$ distribution, 
we define $\Delta(\mathrm{d}N/\mathrm{d}\sin^2{\theta})$ such that~:
\begin{equation}
\label{eqn:DeltadNdsin2th-1}
\Delta(\mathrm{d}N/\mathrm{d}\sin^2{\theta}) = \frac{1}{r}~\bigg(\frac{\mathrm{d}N_{dipole}/\mathrm{d}\sin^2{\theta}-\mathrm{d}N_{iso}/\mathrm{d}\sin^2{\theta}}{\mathrm{d}N_{iso}/\mathrm{d}\sin^2{\theta}}\bigg).
\end{equation}
Once multiplied by the dipole amplitude $r$, $\Delta(\mathrm{d}N/\mathrm{d}\sin^2{\theta})$ gives 
directly the relative changes in the $\mathrm{d}N/\mathrm{d}\sin^2{\theta}$ distribution with respect 
to isotropy. Carrying out integrations over $\varphi$ and $\alpha^0$ yields to~:
\begin{equation}
\label{eqn:DeltadNdsin2th-2}
\Delta(\mathrm{d}N/\mathrm{d}\sin^2{\theta}) = \frac{N_{0,dipole}}{N_{0,iso}}~\sin{\ell_\mathrm{site}}\sin{\delta_d}\cos{\theta},
\end{equation}
where both intensity normalisations $N_{0,iso}$ and $N_{0,dipole}$ are tuned to guarantee the
same number of events observed in the covered region of the sky for each underlying angular 
distribution. This result is shown in the left panel of Fig.~\ref{fig:localangles}, for the latitude 
$\ell_{\mathrm{site}}=-35.2^\circ$ of the Pierre Auger Observatory and for different dipole directions.
Within the zenithal range $[0^\circ,55^\circ]$ considered in this article, the relative changes - maximal for 
$\delta_d=\pm90^\circ$ - amount at most to $\simeq\pm15\%$. So, even for an amplitude $r$ as
large as 10\%, the relative changes in $\mathrm{d}N/\mathrm{d}\sin^2{\theta}$ would be within 
$\simeq\pm1.5\%$, variation which - given the available statistics - is sufficiently low to be considered 
as negligible. Besides, the same calculation applied to the case of a symmetric quadrupolar anisotropy 
shows that the variation of $\Delta(\mathrm{d}N/\mathrm{d}\sin^2{\theta})$ is less than $\simeq 0.1\%$, 
thus being negligible. Consequently, the distribution in $\mathrm{d}N/\mathrm{d}\sin^2{\theta}$ 
can be considered at first order as \emph{insensitive} to large scale anisotropies, so that any significant 
deviation from a uniform distribution provides an empirical measurement of the zenithal dependence 
of the detection efficiency.

It is worth noting that the azimuthal distribution averaged over time is, on the other hand, sensitive 
to large scale anisotropies. Repeating the same calculation and integrating now over $\theta$ 
(in this example between 0 and 60$^\circ$) and 
$\alpha^0$ yields the $\Delta(\mathrm{d}N/\mathrm{d}\varphi)$ relative changes~:
\begin{equation}
\label{eqn:DeltadNdphi}
\Delta(\mathrm{d}N/\mathrm{d}\varphi) = \frac{N_{0,dipole}}{N_{0,iso}}~\frac{\sin{\delta_d}\cos{\ell_\mathrm{site}}}{24}~\bigg(7\tan{\ell_\mathrm{site}}+3\sqrt{3}\sin{\varphi}\bigg).
\end{equation}
This function is shown in the right panel of Fig.~\ref{fig:localangles}, for $\delta_d=90^\circ$ (dashed
line) and $\delta_d=-90^\circ$ (dotted line). The amplitude of the dipole wave is now $\simeq 0.5$. 
As well, the influence of a quadrupole on $\Delta(\mathrm{d}N/\mathrm{d}\varphi)$ is illustrated 
by the dashed-dotted line (oblate symmetric quadrupole in this example). Since, at the Earth latitude of 
the Pierre Auger Observatory, any genuine large scale pattern which depends on the declination 
translates into azimuthal modulations of the event rate \emph{similar} to the ones induced by 
experimental effects, it is thus mandatory to model accurately the dependence on azimuth of the 
detection efficiency for disentangling local from celestial effects. 

\section*{Appendix B~: Modulation of the detection efficiency induced by a tilted array}
\label{appendixB}

To estimate the modulation of the detection efficiency induced by a tilted array, we consider
here that in the absence of tilt, the corresponding detection efficiency function $\epsilon_{\mathrm{notilt}}$  
depends \textit{only} on the energy and the zenith angle and can be parameterised in a good 
approximation as~:
\begin{equation}
\epsilon_{\mathrm{notilt}}(E,\theta)=\frac{E^3}{E^3+E_{0.5}^3(\theta)}.
\end{equation}
$E_{0.5}(\theta)$ is the zenithal-dependent energy at which $\epsilon_{\mathrm{notilt}}(E,\theta)=0.5$.
In case of a tilted array, this parameter depends also on the azimuth angle, which is then the 
source of the azimuthal modulation of the detection efficiency. To understand this, it is useful 
to consider for any given shower with parameters $(E,\theta,\varphi)$ the circle in the shower
plane corresponding to the region in which a signal $S$ larger than some specified threshold 
value $S_0$ is expected. Let $r_0(\zeta)$ denote the radius of this circle, $\zeta$ being the
tilt angle of the SD array. The detection efficiency,
and hence also the parameter $E_{0.5}$, is ultimately a function of the average number of
detectors contained in the projection of this circle into the ground, given by~:
\begin{equation}
\left<n_{\mathrm{det}}\right>(S>S_0)\propto\frac{r_0^2}{h^2 |\mathbf{n_\perp} \cdot \mathbf{n} |},
\end{equation}
where $h=1.5~$km is the nominal separation between surface detectors. The radii $r_0(\zeta)$
obtained with the tilted array leading to the same value of $\left<n_\mathrm{det}\right>$ can be related 
to $r_0(\zeta=0)$ through~:
\begin{equation}
r_0^2(\zeta)=r_0^2(\zeta=0)\frac{| \mathbf{n_\perp}\cdot \mathbf{n} |}{\cos{\theta}}.
\end{equation}
Hence, we can obtain the relation between the energies $E_{0.5}$ with tilt $(E_{0.5}^{\mathrm{tilt}})$ 
and without tilt $(E_{0.5})$ by comparing the cosmic ray energies required to get the value 
$S_0$ at radius $r_0(\zeta)$ and at radius $r_0(\zeta=0)$. Approximating the lateral distribution
function of the signal near the radius $r_0$ as a power law $S(r)\propto Er^{-3}$, we obtain
the following relation~:
\begin{equation}
E_{0.5}^{\mathrm{tilt}}(\theta,\varphi)=E_{0.5}(\theta)\bigg(\frac{r_0(\zeta)}{r_0(\zeta=0)}\bigg)^3\simeq E_{0.5}(\theta)[1+\zeta\tan{\theta}\cos{(\varphi-\varphi_0)}]^3.
\end{equation}
Then, subtracting $\epsilon_{\mathrm{notilt}}$ to $\epsilon_{\mathrm{tilt}}$ leads to Eqn.~\ref{eqn:tilt2}.

\section*{Appendix C~: Determination of upper limits on dipole amplitudes}
\label{appendixC}

To determine upper limits on the dipole amplitudes, Linsley described the procedure to
follow in the case of first harmonic analysis in right ascension~\citep{Linsley1975}. We
adapt here this procedure to the case of the dipolar reconstruction adopted in 
section~\ref{dipole}.

Here, the data set is supposed to have been drawn at random from an underlying 
dipolar distribution characterised by \textbf{d}, whose value is unknown. In the limit of large
number of events, the joint p.d.f. $p_{D_X,D_Y,D_Z}(\overline{d}_x,\overline{d}_y,\overline{d}_z)$ 
can be factorised in terms of three Gaussian distributions $N(\overline{d}_i-d_i,\sigma_i)$~:
\begin{equation}
\label{eqn:pdfdipole}
p_{D_X,D_Y,D_Z}(\overline{d}_x,\overline{d}_y,\overline{d}_z;d_x,d_y,d_z) = N(\overline{d}_x-d_x,\sigma) N(\overline{d}_y-d_y,\sigma) N(\overline{d}_z-d_z,\sigma_z).
\end{equation}
The joint p.d.f. $p_{R,\Delta,A}(\overline{r},\overline{\delta},\overline{\alpha})$ expressing the 
dipole components in spherical coordinates is then obtained by performing the Jacobian
transformation~:
\begin{eqnarray}
\label{eqn:pdfdipole2}
p_{R,\Delta,A}(\overline{r},\overline{\delta},\overline{\alpha};d,\delta_d,\alpha_d) &=& \bigg| \frac{\partial(\overline{d}_x,\overline{d}_y,\overline{d}_z)}{\partial(\overline{r},\overline{\delta},\overline{\alpha})}\bigg|p_{D_X,D_Y,D_Z}(\overline{d}_x(\overline{r},\overline{\delta},\overline{\alpha}),\overline{d}_y(\overline{r},\overline{\delta},\overline{\alpha}),\overline{d}_z(\overline{r},\overline{\delta},\overline{\alpha}))\nonumber \\
&=&\frac{\overline{r}^2\cos{\overline{\delta}}}{(2\pi)^{3/2}\sigma^2\sigma_z}\exp{\bigg[-\frac{(\overline{r}\sin{\overline{\delta}}-d\sin{\delta_d})^2}{2\sigma_z^2}\bigg]}\nonumber\\
&~~~~~~\times&\exp{\bigg[-\frac{(\overline{r}\cos{\overline{\delta}}\cos{\overline{\alpha}}-d\cos{\delta_d}\cos{\alpha_d})^2}{2\sigma^2}\bigg]}\nonumber \\
&~~~~~~\times &\exp{\bigg[-\frac{(\overline{r}\cos{\overline{\delta}}\sin{\overline{\alpha}}-d\cos{\delta_d}\sin{\alpha_d})^2}{2\sigma^2}\bigg]}.
\end{eqnarray}
Each analysed data set having been selected at random from an ensemble in which all possible values
of $\mathbf{d}$ are equally represented, the various $d$, $\delta_d$ and $\alpha_d$ combinations have
relative probability $p_{R,\Delta,A}(\overline{r},\overline{\delta},\overline{\alpha};d,\delta_d,\alpha_d)/p_{R,\Delta,A}(\overline{r},\overline{\delta},\overline{\alpha};d=0)$. This allows us to define the joint p.d.f.
$\tilde{p}_{R,\Delta,A}$ by requiring this ratio to be normalised to unity~:
\begin{eqnarray}
\label{eqn:pdful1}
\tilde{p}_{R,\Delta,A}(\overline{r},\overline{\delta},\overline{\alpha};d,\delta_d,\alpha_d) &=& K(r,\delta)~\exp{\bigg[\frac{\overline{r}d\cos{\overline{\delta}}\cos{\delta_d}\cos{(\overline{\alpha}-\alpha_d)}}{\sigma^2}\bigg]} \nonumber \\
&~~~~~~\times&\exp{\bigg[\frac{\overline{r}d\sin{\overline{\delta}}\sin{\delta_d}}{\sigma^2_z}-\frac{d^2\cos^2{\delta_d}}{2\sigma^2}-\frac{d^2\sin^2{\delta_d}}{2\sigma_z^2}\bigg]},
\end{eqnarray}
where the normalisation reads~:
\begin{eqnarray}
\label{eqn:pdful-norm}
K(r,\delta)&=&2\pi~I_0\bigg(\frac{\overline{r}d\cos{\overline{\delta}}\cos{\delta_d}}{\sigma^2}\bigg)\nonumber \\
&~~~~~~\times&\int~\mathrm{d}d~\mathrm{d}\delta_d~\exp{\bigg[-\frac{d^2\cos^2{\delta_d}}{2\sigma^2}-\frac{d^2\sin^2{\delta_d}}{2\sigma_z^2}+\frac{\overline{r}d\sin{\overline{\delta}}\sin{\delta_d}}{\sigma^2_z}\bigg]}.
\end{eqnarray}
$I_0$ is here the modified Bessel function of the first kind with order 0. Integration of $\tilde{p}_{R,\Delta,A}$
over $\delta_d$ and $\alpha_d$ yields the $\tilde{p}_R$ p.d.f., from which upper limits on $d$ can be obtained
within a confidence level $C.L.$ by inverting the relation~:
\begin{eqnarray}
\label{eqn:pdful2}
\int_{\overline{r}_{data}}^1 \mathrm{d}\overline{r}~\tilde{p}_{R}(\overline{r},\overline{\delta};d^{UL}) =C.L.
\end{eqnarray}
Due to the non-uniform directional exposure in declination, the resulting upper limits actually depend on
the declination through the dependence of $\tilde{p}_R$ on $\overline{\delta}$. In practice, this dependence
is small, which is why we presented in section~\ref{discussion} upper limits \textit{averaged} over the
declination.

\section*{Appendix D~: Determination of upper limits on quadrupole amplitudes}
\label{appendixD}

To determine upper limits on quadrupole amplitudes, we rely on Monte-Carlo simulations. 
For each possible amplitude $\lambda_+$ ($\beta$), we estimate the p.d.f. 
$p_{\Lambda_+}(\overline{\lambda}_+;\lambda_+)$ ($p_{B}(\overline{\beta};\beta)$) 
with a given number of events $N$ and a given exposure $\tilde{\omega}$. The amplitude 
$\lambda_+^{UL}$ such that $\int_{\overline{\lambda}_{+,data}}^\infty \mathrm{d}\overline{\lambda}_+~\tilde{p}_{\Lambda}(\overline{\lambda}_+;\lambda_+^{UL}) =C.L.$
is a relevant upper limit (and respectively for $\beta^{UL}$).

Alternatively to the previous procedure used to derive upper limits on dipole amplitudes, this
procedure can lead to upper limits tighter than the upper bounds for isotropy 
$\overline{\lambda}_{+,99}$ when the measured values of $\overline{\lambda}_{+,data}$ 
are smaller than the expected average for isotropy. To cope with this 
undesired behaviour, the upper limits presented in section~\ref{discussion} are 
defined as $\mathrm{max}(\overline{\lambda}_{+,99},\lambda_+^{UL})$.

\end{document}